\documentclass[pre,onecolumn,reprint,aps,superscriptaddress,longbibliography]{revtex4-1}
\usepackage{feynmf}
\usepackage{graphicx}%
\usepackage{dcolumn}%
\usepackage{bm}%
\usepackage{color}
\usepackage{multirow}
\usepackage[normalem]{ulem}
\usepackage{amsmath}
\usepackage{enumerate}
\usepackage{amsfonts}
\usepackage{epsfig}
\usepackage{graphicx}
\usepackage{cancel}
\usepackage[caption=false]{subfig}
\usepackage{floatrow}
\usepackage{subfig}
\newcommand{\be}{\begin{equation}}
\newcommand{\ee}{\end{equation}}
\newcommand{\bea}{\begin{eqnarray}}
\newcommand{\eea}{\end{eqnarray}}

\definecolor{green}{rgb}{0.0, 0.44, 0.0}
\definecolor{red}{rgb}{1.0, 0.13, 0.32}
\definecolor{blue}{rgb}{0.06, 0.2, 0.65}
\definecolor{darkgreen}{rgb}{0,0.5,0}
\definecolor{darkblue}{rgb}{0,0,0.6}
\definecolor{purple}{rgb}{0.4,.2,0.7}
\definecolor{magenta}{rgb}{1.0,0.0,1.0}
\usepackage[colorlinks=true,citecolor=darkgreen,linkcolor=purple,urlcolor=purple]{hyperref}

\def\le{\left}
\def\ri{\right}
\def\T{\mathcal{X}}

\def\M{\mathrm{M}}

\def\gbare{g_{\mathrm{bare}}}
\def\Pa{\hat{a}}
\def\Pb{\hat{b}}

\begin{document}

\title{Nontrivial critical fixed point for replica-symmetry-breaking transitions} 

\author{Patrick Charbonneau} 
\affiliation{Department of Chemistry, Duke University, Durham,
North Carolina 27708, USA}
\affiliation{Department of Physics, Duke University, Durham,
North Carolina 27708, USA}

\author{Sho Yaida}
\email{sho.yaida@duke.edu}
\affiliation{Department of Chemistry, Duke University, Durham, 
North Carolina 27708, USA}

\begin{abstract}
The transformation of the free-energy landscape from smooth to hierarchical is one of the richest features of mean-field disordered systems. A well-studied example is the de Almeida-Thouless transition for spin glasses in a magnetic field, and a similar phenomenon--the Gardner transition--has recently been predicted for structural glasses. The existence of these replica-symmetry-breaking phase transitions has, however, long been questioned below their upper critical dimension, $d_{\rm u}=6$. Here, we obtain evidence for the existence of these transitions in $d<d_{\rm u}$ using a two-loop calculation. Because the critical fixed point is found in the strong-coupling regime, we corroborate the result by resumming the perturbative series with inputs from a three-loop calculation and an analysis of its large-order behavior. Our study offers a resolution of the long-lasting controversy surrounding phase transitions in finite-dimensional disordered systems.
\end{abstract}

\maketitle

\begin{figure*}[t]
\centerline{
\hspace{-1.0in}
\subfloat[One-loop RG, $d<6$\ \ \ \ \ \ \ \ ]{\includegraphics[width=0.315\textwidth]{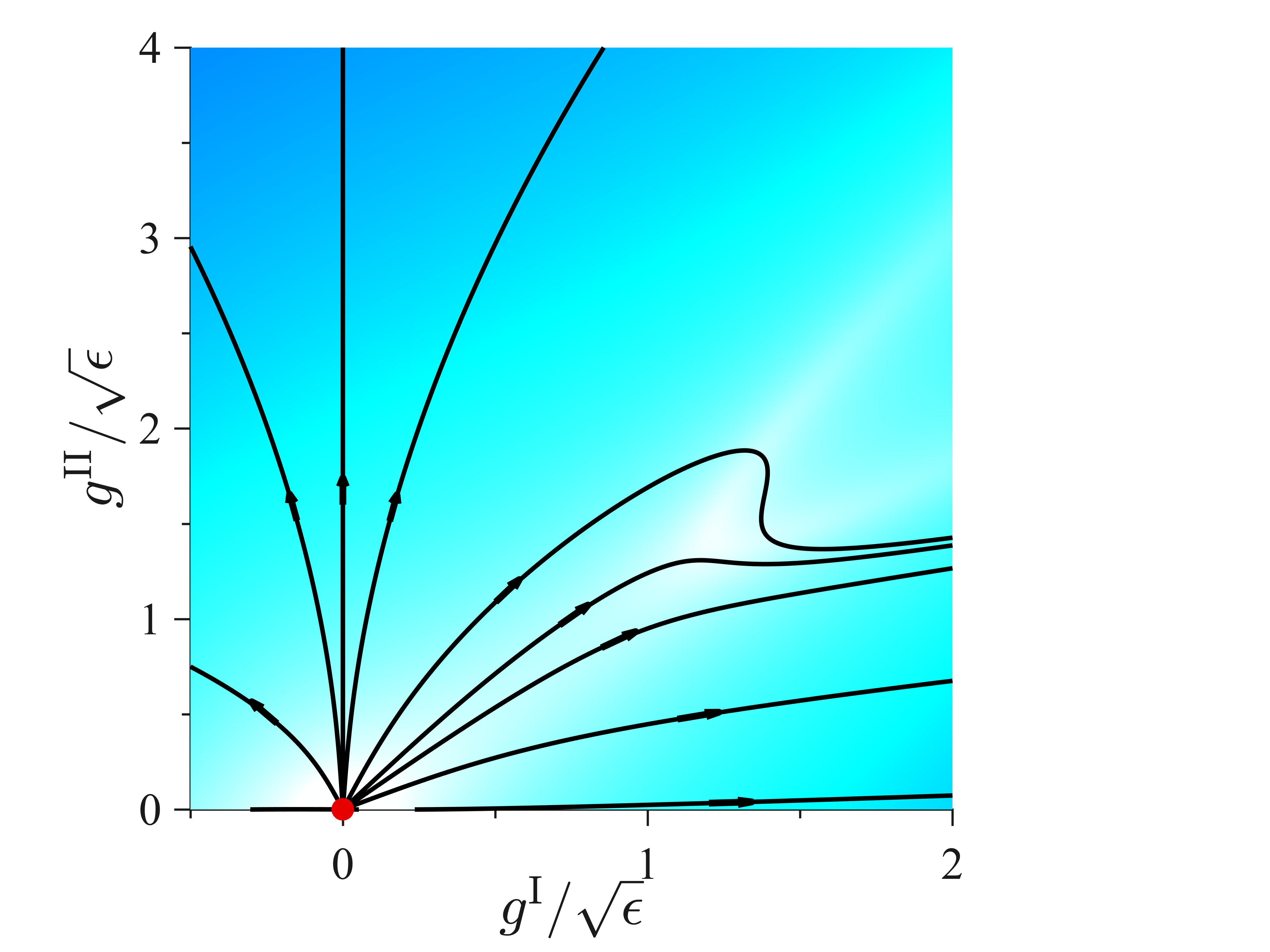}}\quad%
\hspace{-0.7in}
\subfloat[Two-loop RG, $d=3$\ \ \ \ \ \ \ \ ]{\includegraphics[width=0.315\textwidth]{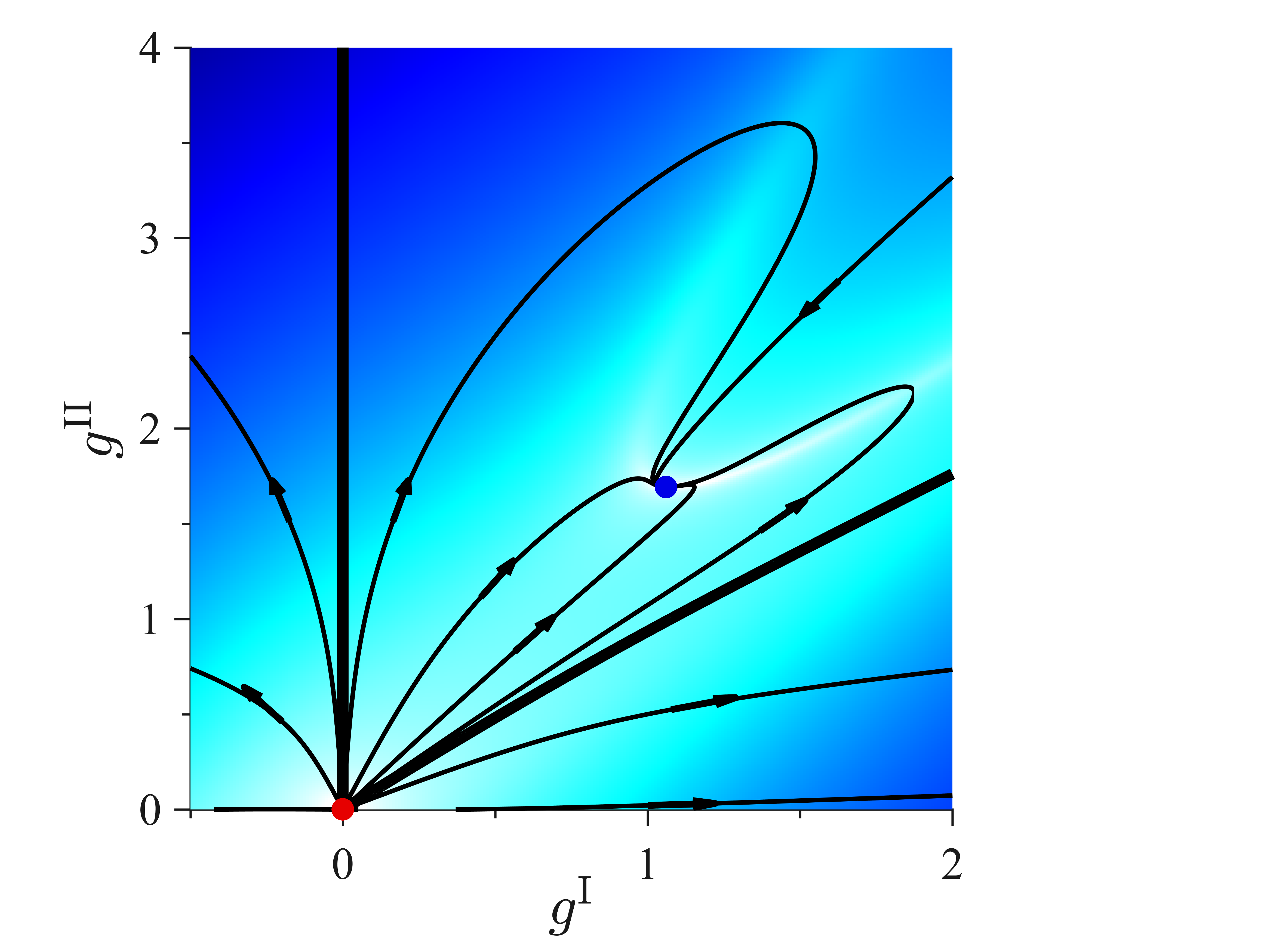}}\quad%
\hspace{-0.7in}
\subfloat[Two-loop RG, $d=5$\ \ \ \ \ \ \ \ ]{\includegraphics[width=0.315\textwidth]{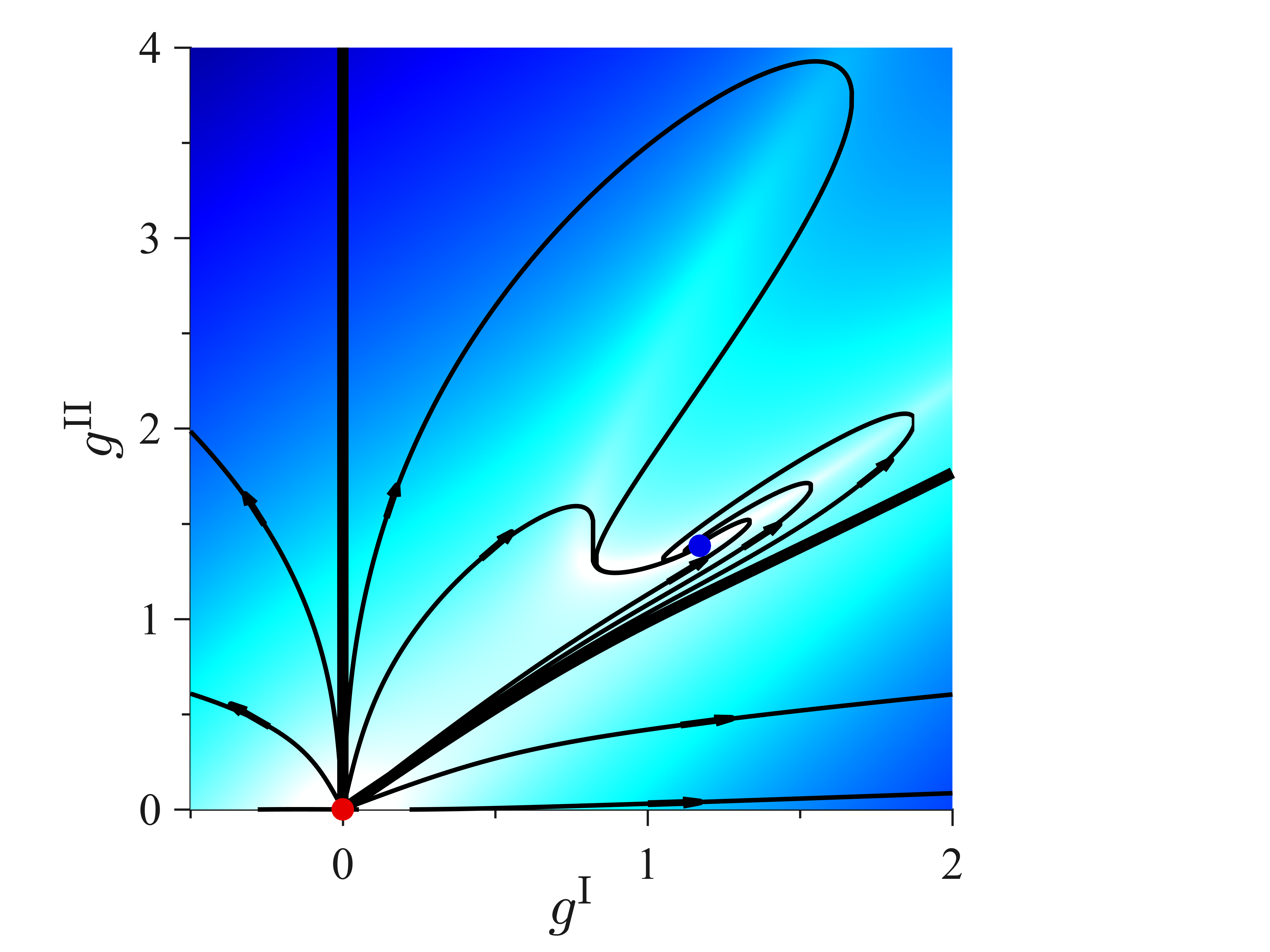}}\quad%
\hspace{-0.7in}
\subfloat[Two-loop RG, $d=5.5$\ \ \ \ \ \ \ \ ]{\includegraphics[width=0.315\textwidth]{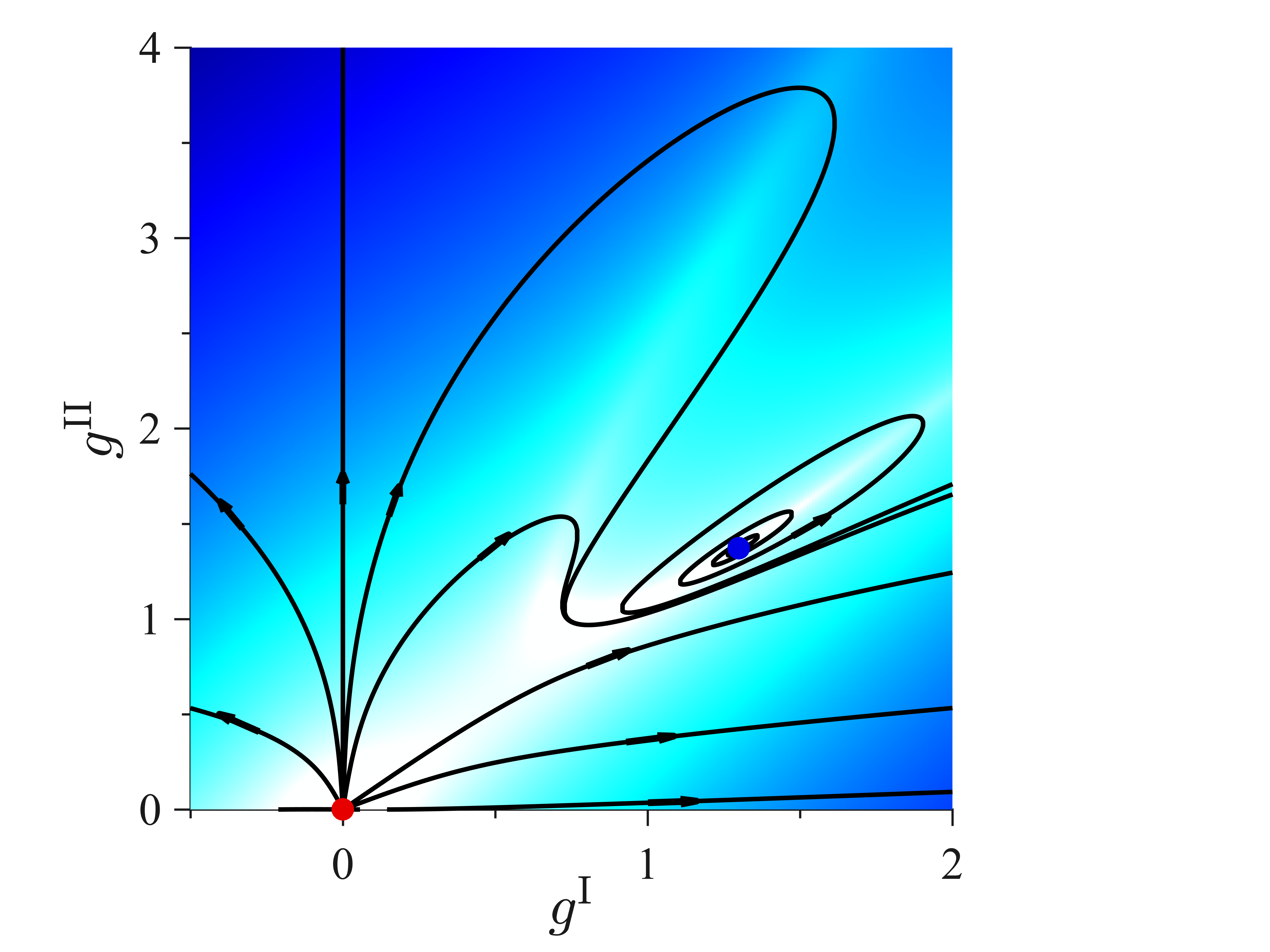}}
\hspace{-1in}
}
\vspace{-1.86in}
\centerline{\hspace{+8.5in}
\includegraphics[width=0.284\textwidth]{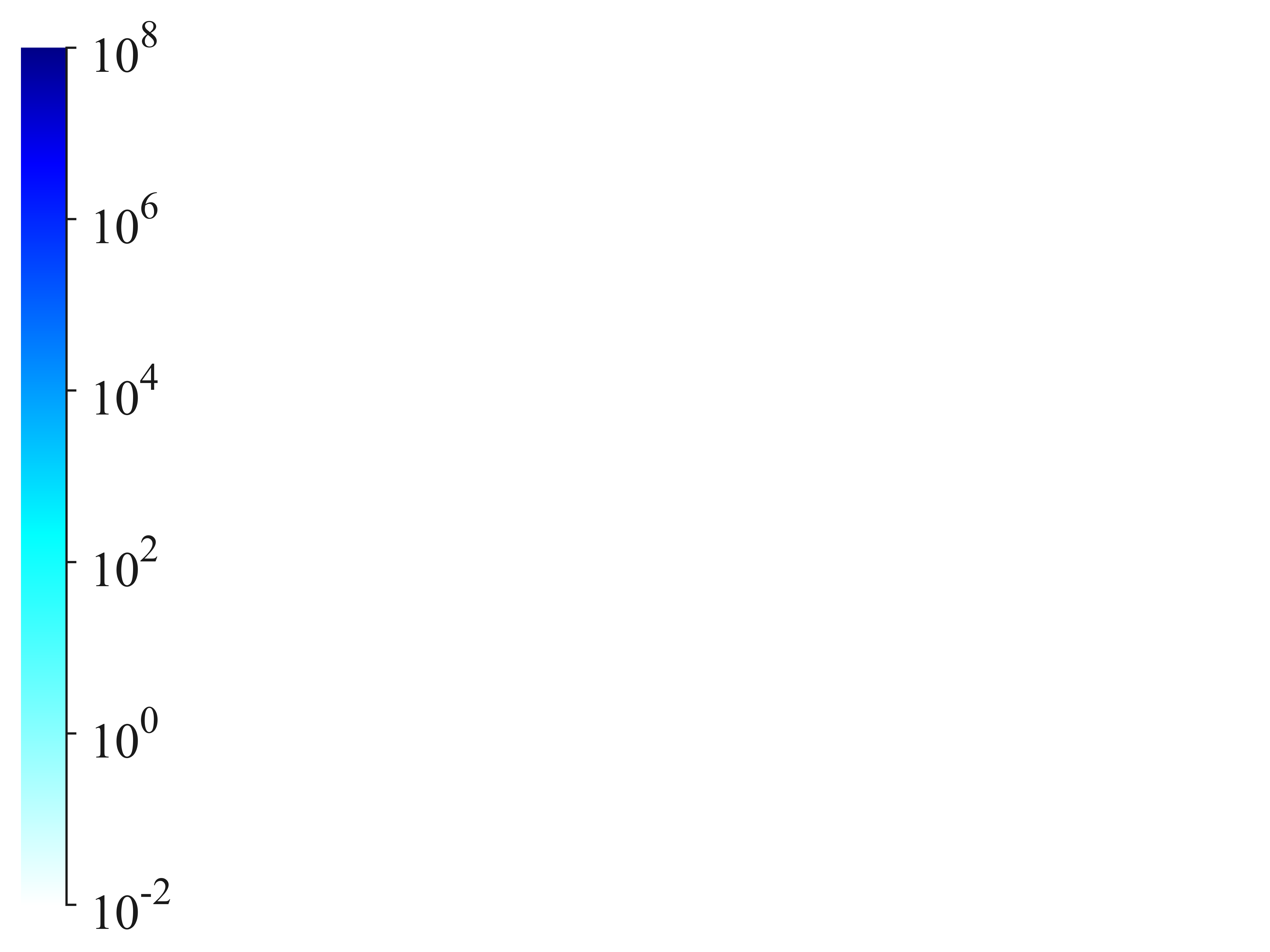}
}
\vspace{+0.25in}
\caption{RG flows in the space of couplings for (a) the one-loop calculation in $d<6$ and the two-loop calculation in (b) $d=3$, (c) $d=5$ and (d) $d=5.5$. Arrows denote flow toward longer length scales;
background shading denotes the intensity of the flow quantified by $\le(\beta^{\mathrm{I}}\ri)^2+\le(\beta^{\mathrm{II}}\ri)^2$, and normalized by $\epsilon^{-3}$ in (a).
The Gaussian fixed point (red dot) is unstable for $d<6$.
In (b) and (c) a nontrivial fixed point (blue dot) is stable and lies at strong couplings. Its basin of attraction is delineated by two thick lines: one precisely along $g^{\mathrm{I}}=0$ and the other approximately along $g^{\mathrm{I}}\approx g^{\mathrm{II}}$. Outside this basin, the flow runs toward infinity, which is often characteristic of discontinuous transitions.
Note that for $d=5$, the flow spirals into the nontrivial fixed point, while for $d=5.5$ both fixed points are unstable.}
\label{RGflow}
\end{figure*}

\emph{Introduction--} Spontaneous symmetry breaking can dramatically change material properties. Breaking translational symmetry turns liquids into crystalline solids,
%breaking spin rotational symmetry makes magnetism permanent, 
breaking gauged phase symmetry gives rise to superconductivity, and breaking non-Abelian gauged symmetry endows elementary particles with mass.
In a host of disordered models, a symmetry of the most peculiar type can break. Upon cooling, the mean-field free-energy of these systems develops a finite complexity, \textit{i.e.}, the number of metastable states grows exponentially with system size. The similitude between copies (replicas) of the system then depends on whether or not they belong to a same cluster of metastable states. In particular, right at the transition point, each replica of the system is on the brink of falling into one cluster or another, resulting in critical fluctuations of the similarity between uncoupled copies. Remarkably, such replica symmetry breaking (RSB) accounts for the emergence of glassiness in mean-field models ranging from liquids to optimization problems and neural networks~\cite{MPV87}.  Mean-field criticality, however, bears the seed of its own destruction.  Below the upper critical dimension, $d_{\rm u}$, violent critical fluctuations challenge the very validity of the approximation within which they were conceived. The existence of a continuous transition into an RSB phase in dimensions $d<d_{\rm u}$ is thus not a foregone conclusion, and its fate in disordered systems remains hotly debated~\cite{Parisi12}.

An illustrious example of this dispute centers around mean-field models of spins with quenched impurities in an external magnetic field, known to exhibit a de Almeida-Thouless (dAT) transition~\cite{AT78}. This transition accompanies the emergence of continuous RSB with a hierarchically rough landscape, which eventually becomes fractal in the low-temperature limit~\cite{CKPUZ14}. Its upper critical dimension, $d_{\rm u}=6$~\cite{HLC76}, however, is well above three and the existence of the transition in real physical systems has long been questioned~\cite{McMillan84,FH86,HF87,FH88,BM87,NS00,NS03,MB11}.
Recent advances in the mean-field description of structural glasses have unveiled a new facet of this problem.
Solid glasses are also predicted to undergo a critical RSB transition--known as a Gardner transition--upon cooling, compressing or shearing~\cite{CKPUZ14,RUYZ15,BU16}. A growing body of evidence further relates the Gardner transition to the anomalous behavior of amorphous solids compared to their crystalline counterpart, and to a nontrivial critical scaling upon approaching the jamming limit~\cite{CKPUZ14,DLBW14,DLW15,CCPZ15,CCPPZ16,SD16}. The question of whether dAT and Gardner transitions survive finite-dimensional fluctuations has thus gained renewed impetus.

The impact of fluctuations on RSB transitions was first examined using the perturbative renormalization group (RG) approach that proved so successful for Ising and other universality classes. A loop-expansion of the field theory appropriate for the dAT and Gardner universality class, however, finds that the critical fixed point is absent to lowest, one-loop order for $d<d_{\rm u}$~\cite{BR80,PTD02,BU15}. This has led many to conclude that such transitions then either become discontinuous or simply vanish.  Yet the lack of dimensional robustness is challenged by numerical evidence supporting the existence of a critical dAT transition in $d=4$~\cite{JANUS12,LPRR09} and of a critical Gardner transition in $d=3$~\cite{BCJPSZ16,footnote_statefollowing}. An alternate interpretation is that the critical fixed point resides in the strong-coupling regime of the field theory for all $d$, thus preventing a one-loop calculation from identifying it. A precedent is the Caswell-Banks-Zaks (CBZ) fixed point of the non-Abelian gauge theory for elementary particles in $3+1$ space-time dimensions~\cite{Caswell74}. This fixed point is missed at one-loop order but captured at two-loop order.
%The reason can be gleaned from the distance-dependence of the interaction strength between elementary particles, \textit{i.e.}, the coupling constant $g$. The rate of change in $g$ is governed by a $\beta$-function \be \beta\equiv\mu\frac{\partial g}{\partial \mu}=\beta_{1\mathrm{-loop}} g^3+\beta_{2\mathrm{-loop}} g^5+O(g^7) , \ee where $\mu$ is the inverse-length (or energy) scale of interest. Retaining only terms proportional to $\beta_{1\mathrm{-loop}}$ results in two cases: (i) if $\beta_{1\mathrm{-loop}}>0$, the coupling constant approaches zero at long distances (\textit{i.e.}, low energies), which is the screening effect in quantum electrodynamics; (ii) if $\beta_{1\mathrm{-loop}}<0$, the interaction strength grows indefinitely as a pair of particles are pulled apart, which is the anti-screening effect that results in quark confinement in quantum chromodynamics~\cite{GW73,P73} ($\beta_{1\mathrm{-loop}}=0$ is marginal). In both cases, the RG flow of coupling only stops at the Gaussian $g=0$ fixed point, which is infrared-stable for (i) and unstable for (ii). At two-loop order, however, a stable fixed point can also exist for case (ii) if $\beta_{2\mathrm{-loop}}>0$ (see Ref.~\cite[Fig.~1]{Caswell74}).
Although the CBZ fixed point generically lies in the strong-coupling regime, which falls beyond the designed range of a perturbative calculation, its existence has been corroborated by adiabatically connecting it to a perturbative fixed point~\cite{BZ82}, supported by lattice simulations even in the strong-coupling regime~\cite{Debbio11}, and established beyond reasonable doubt in supersymmetric theories~\cite{Seiberg95,Strassler05}.  Two-loop calculations may thus find fixed points that are missed by one-loop analysis, but additional lines of evidence are then needed to confirm the result. 

In this letter, we present field-theoretic calculations that capture the physics of both dAT transitions in spin glasses and Gardner transitions in structural glasses. Like for the CBZ fixed point, our two-loop calculation identifies a critical fixed point for $d<d_{\mathrm{u}}$ that is missed by the one-loop RG flow. Resummation of the perturbative series at three-loop order supplemented by an analysis of its large-order behavior further supports the robustness of this critical fixed point for the dAT-Gardner universality class.

\emph{Field-Theory Setup--} The finite-dimensional generalization of the mean-field Edwards-Anderson order parameter for glasses is the replicated overlap field, $q_{ab}\le(\mathbf{x}\ri)$. This field characterizes the similarity at positions $\mathbf{x}$ between pairs of distinct replicated configurations through an $n$-by-$n$ symmetric matrix with a null diagonal; the zero replica limit, $n\rightarrow0$, is taken at the end of the calculations in order to properly average over disorder (see Appendix~\ref{Lagrangian}). In general, field fluctuations can be subdivided into longitudinal, anomalous, and replicon modes~\cite{BM79,NishimoriBook}. At dAT and Gardner transitions only replicon modes become critical (massless); the other two remain short-ranged (massive) and can thus be neglected at long distances. We henceforth only focus on the replicon field, $\phi_{ab}\le(\mathbf{x}\ri)$, defined by the condition $\sum_{b=1}^{n}\phi_{ab}=0$ for all $a=1,\ldots,n$, thus leaving $n(n-3)/2$ degrees of freedom.

In order to investigate the putative critical point, we seek infrared-stable fixed points of the RG flow within the critical surface on which the replicon field remains massless. Within this surface, the field theory is governed by the bare action, $S=\int \mathrm{d}\mathbf{x} \mathcal{L}$, with~\cite{footnote_coupling}
\bea\label{eq:action}
\mathcal{L}&=&\frac{1}{2}\sum_{a,b=1}^{n}\le(\nabla\phi_{ab}\ri)^2\\
&&-\frac{1}{3!}\le[g_{\rm bare}^{\rm I}\sum_{a,b=1}^n\phi_{ab}^3+g_{\rm bare}^{\rm II}\sum_{a,b,c=1}^n\phi_{ab}\phi_{bc}\phi_{ca}\ri]\, ,\nonumber
\eea
which is the most generic cubic action for replicon modes~\cite{PTD02}. The effective description of the system then depends on the energy scale, $\mu$, probed. This dependence is encoded in the RG flow of dimensionless couplings, $g^{\T}(\mu)$ with $\T\in\{\mathrm{I}, \mathrm{II}\}$, that are related to bare couplings, $g_{\rm bare}^{\T}$, in Eq.~\eqref{eq:action} (see Appendix~\ref{2beta}). The flow is governed by $\beta^{\T}\equiv\mu \partial g^{\T}/\partial \mu$, and stops at fixed points whereat $\beta^{\mathrm{I}}\le(g_{\star}^{\mathrm{I}},g_{\star}^{\mathrm{II}}\ri)=\beta^{\mathrm{II}}\le(g_{\star}^{\mathrm{I}},g_{\star}^{\mathrm{II}}\ri)=0\, $. Note that for all $d$ a Gaussian fixed point with $g_{\star}^{\mathrm{I}}=g_{\star}^{\mathrm{II}}=0$ exists, but it is stable only for $d>d_{\rm u}$~\cite{BR80}.

\begin{figure*}[t]
\centerline{
\sidesubfloat[]{\hspace{-0.065in}\includegraphics[width=0.33\textwidth]{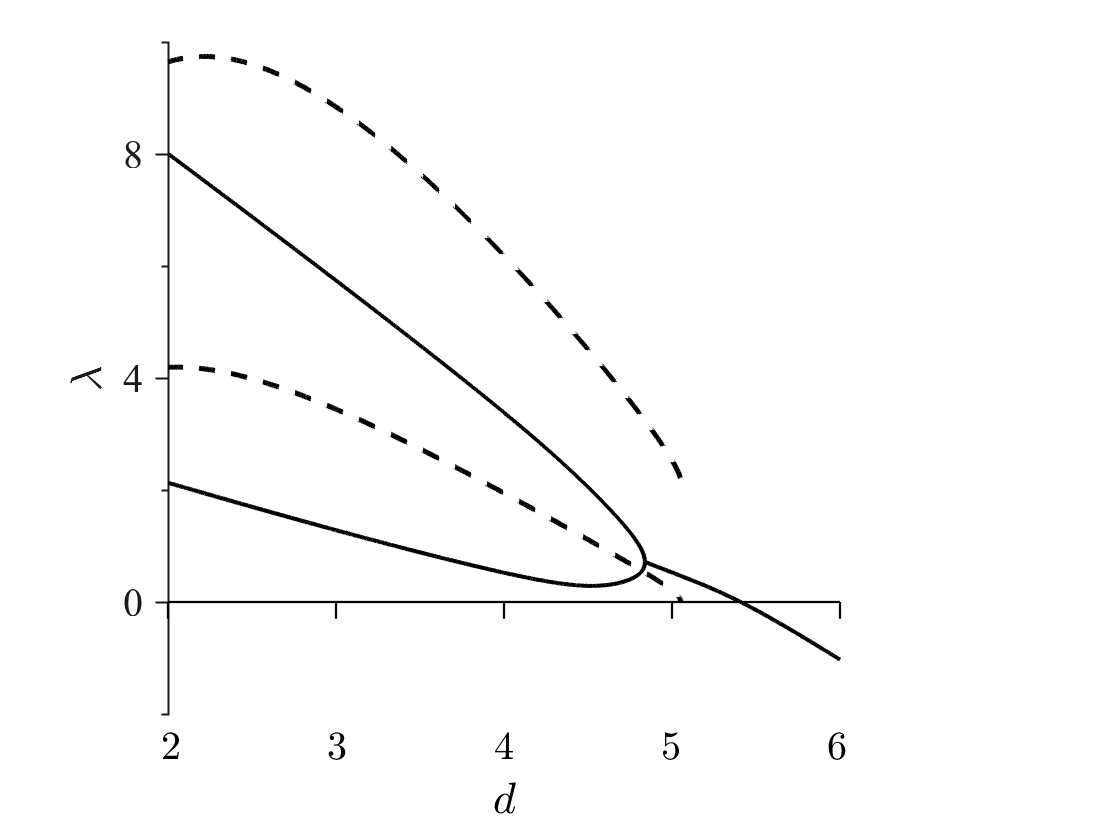}}\quad%
\hspace{-0.3in}
\sidesubfloat[]{\hspace{-0.065in}\includegraphics[width=0.33\textwidth]{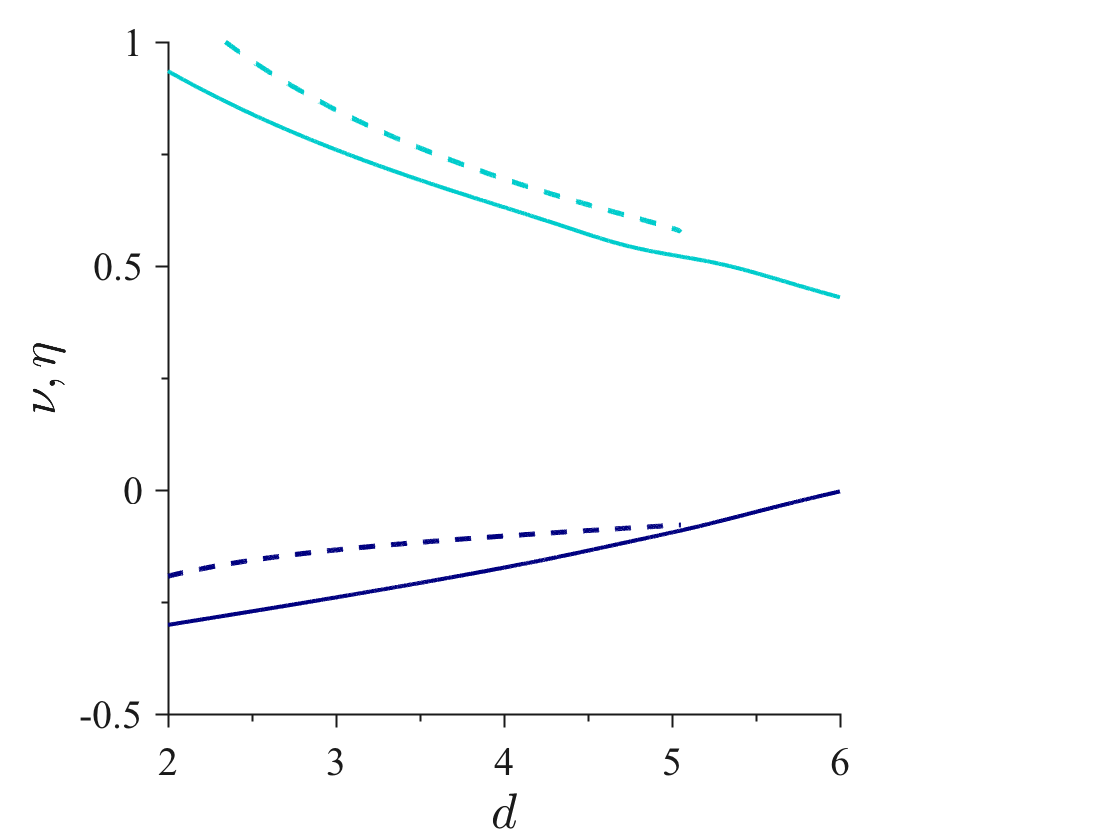}}\quad%
\hspace{-0.3in}
\sidesubfloat[]{\hspace{-0.065in}\includegraphics[width=0.33\textwidth]{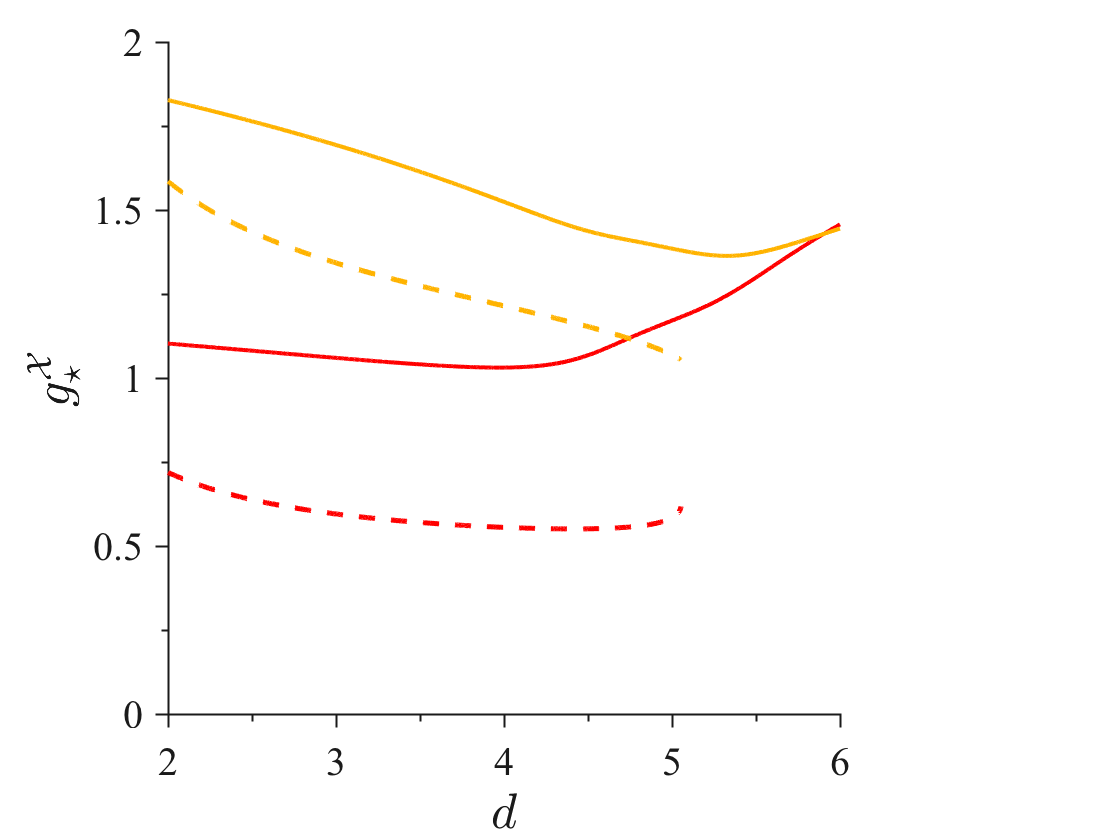}}
}
\caption{Critical parameters at the nontrivial fixed point derived within two-loop (solid lines) and Borel resummation (dashed lines) RG schemes as functions of the spatial dimension $d$.
(a) Real parts of the stability exponents around the fixed point within the critical surface, $\lambda_1$ and $\lambda_2$.
(b) Critical exponents, $\nu$ (cyan) and $\eta$ (navy-blue).
(c) Fixed-point values of running couplings, $g^{\mathrm{I}}$ (red) and $g^{\mathrm{II}}$ (orange).
At two-loop order, the nature of the fixed point changes at $d_{\rm s}\approx 4.84$ and $d_{0}\approx5.41$. The two stability exponents merge at $d=d_{\rm s}$, at which point they acquire imaginary parts, hence the flow spirals into the (stable) fixed point [Fig.~\ref{RGflow}(c)], while for $d>d_{0}$ the real part of these eigenvalues becomes negative and the flow spirals out of the (unstable) fixed point [Fig.~\ref{RGflow}(d)]. Upon inclusion of higher-loop corrections, Borel resummation indicates that the fixed point is robustly stable for $d\lesssim 5.05$ but does not exhibit any spiraling flow.}
\label{NFPinfo}
\end{figure*}

\emph{Two-loop RG--} Inspired by the CBZ fixed point, we compute the $\beta$-functions to two-loop order for the replica field theory in Eq.~\eqref{eq:action}, using the dimensional regularization scheme~\cite{HV72,H73a,H73b,Caswell74,Amit76,BKM81,Gracey15} (see Appendix~\ref{2beta}). As expected~\cite{BR80,PTD02,BU15}, no stable fixed point can be found at one-loop order for $d<6$ [Fig.~\ref{RGflow}(a)]. For $d<d_{0}\approx5.41$, however, the two-loop RG flow locates a stable fixed point with a finite basin of attraction [Figs.~\ref{RGflow}(b) and (c)]. A system lying within this basin eventually approaches the fixed point upon rescaling and is thus critical. By contrast, a system that remains outside the basin cannot continuously transition into an RSB phase, and may instead exhibit a discontinuous transition.  Remarkably, the boundary of the basin is closely approximated by the tree-level condition for a critical transition into a RSB phase, \textit{i.e.}, $1<g^{\mathrm{II}}/g^{\mathrm{I}}<\infty$~\cite{FPR96}.

The eigenvalues, $\lambda_{1}$ and $\lambda_2$, of
\be
\left[ {\begin{array}{cc}
  \frac{\partial\beta^{\mathrm{I}}}{\partial g^{\mathrm{I}}} &  \frac{\partial\beta^{\mathrm{I}}}{\partial g^{\mathrm{II}}} \\
    \frac{\partial\beta^{\mathrm{II}}}{\partial g^{\mathrm{I}}} & \frac{\partial\beta^{\mathrm{II}}}{\partial g^{\mathrm{II}}} \\
  \end{array} } \right]\Bigg|_{(g^{\mathrm{I}},g^{\mathrm{II}})=(g_{\star}^{\mathrm{I}},g_{\star}^{\mathrm{II}})}\, 
\ee
give the stability exponents that control subleading corrections from irrelevant deformations near the critical point. Figure~\ref{NFPinfo}(a) indicates that these exponents acquire an imaginary component for $d>d_{\rm s}\approx4.84$, hence the RG flow then spirals toward the fixed point [Fig.~\ref{RGflow}(c)]. As has been observed in other disordered systems~\cite{Aharony75,CL77,WH83}, such complex exponents can emerge from the nonunitarity of the replica field theory, and give rise to an oscillatory decay of the appropriate correlation functions in the critical region. Conformality gets lost with the change in the spiral direction at $d=d_{0}$ and no stable fixed point can be found for $d\in(d_0,d_{\rm u})$ [Fig.~\ref{RGflow}(d)]. In the absence of additional nontrivial fixed points with which to collide~\cite{KLSS09}, this scenario provides a natural mechanism for exchanging dominance between the Gaussian and the genuinely nonperturbative fixed points as one goes from $d>d_{\rm u}$ down to physical dimensions.

We also compute the critical exponents, $\nu$ and $\eta$, that govern the divergence of the correlation length and the decay of two-point correlation functions at the critical point, respectively [Fig.~\ref{NFPinfo}(b)]. The former is obtained from the relevant deformation by the quadratic coupling that drives the system away from the critical surface (see Appendix~\ref{CE}). Estimates of $\nu$ and $\eta$ agree qualitatively with the trend observed in $d=4$ simulations~\cite{JANUS12}; $\eta$ is negative and $\nu$ is larger than its mean-field value, $\nu_{\rm MF}=\frac{1}{2}$.

\emph{Resummation--} Because the critical couplings are of order unity for all $d<d_{\rm u}$ [Fig.~\ref{NFPinfo}(c)], resummation is needed to assess the existence of the fixed point. (Without a careful resummation, even the $d\leq3$ Wilson-Fisher fixed point for the Ising universality class disappears~\cite{BNGM76}.) A field-theoretic perturbative series is indeed generically not convergent but rather asymptotic.  More precisely, a formal series in terms of the coupling constant,
\be\label{asymptotic}
f(g^2)=\sum_{k} f_k g^{2k}\, ,
\ee
typically has coefficients that exhibit a factorial growth, \textit{i.e.}, $f_k\sim k!\, (-1/A)^k$, with a large-order constant $A$ given in terms of the saddle-point action~\cite{Lipatov77,BLZ77}. Although a truncation to the first couple of terms may yield a good approximation in the weak-coupling regime, the series itself is not mathematically well defined.
%If we take the bare perturbative expression for the $\beta$-function~\cite{BNGM76}, the unique nontrivial fixed point is identified at one-, three-, and five-loop orders. However, at two-, four-, and six-loop orders, it comes with an additional spurious fixed point near the upper critical dimension and disappears (through collision) by the time we continue down to $d=3$~\cite{footnote_spurious}. Only through resummation, we can unequivocally retain the Wilson-Fisher fixed point at every order and simultaneously attain triumphant quantitative success at strong coupling though perturbative route.

Borel resummation is the most common scheme used to give epistemological traction to a fixed point. The approach starts from the observation that a Borel transform, $\tilde{f}_{\rm B}(g^2)\equiv\sum_{k} \frac{f_k}{k!} g^{2k}$,
has a finite radius of convergence, $|A|$.
Using the identity $k!=\int_{0}^{\infty} \mathrm{d}t e^{-t} t^k$ the original series [Eq.~\eqref{asymptotic}] can formally be expressed as $
f(g^2)=\int_{0}^{\infty}\mathrm{d}t e^{-t}\tilde{f}_{\rm B}(tg^2).$
The analytic continuation of the Borel transform onto the whole positive axis then unambiguously defines the function $f$. There is typically no problem to this analytic continuation when $A>0$, hence the series is then deemed Borel-summable.

In order to adapt the above scheme to a replica field theory with two cubic couplings, we define $\le(g^{\mathrm{I}},g^{\mathrm{II}}\ri)\equiv g(\cos\theta,\sin\theta)$ and regroup the double series, with the power of $g^2$ counting loop order: 
\bea
f\le(g^{\mathrm{I}},g^{\mathrm{II}}\ri)&=&\sum_{k_1,k_2=0; k_1+k_2=\mathrm{even}}^{\infty}f_{k_1,k_2}\le(g^{\mathrm{I}}\ri)^{k_1}\le(g^{\mathrm{II}}\ri)^{k_2}\\
&=&\sum_{k=0}^{\infty}g^{2k}\le[\sum_{k_1=0}^{2k}f_{k_1,2k-k_1}\le(\cos\theta\ri)^{k_1}\le(\sin\theta\ri)^{2k-k_1}\ri]\, \nonumber\\
&\equiv&\sum_{k=0}^{\infty}f_k\le(\theta\ri)g^{2k} \, .\nonumber
\eea
The Borel-summability of the series is then governed by the angle-dependent large-order behavior $f_k(\theta)\sim k!\, [-1/A(\theta)]^k$.
Consequently, as has been observed for the Abelian gauge theory with background fields~\cite{Dunne04}, Borel-summability depends on the ratio of two couplings, as encoded in the saddle-point solution to the classical equations of motion for replicons~\cite{footnote_Lagrange}.

Among nontrivial saddles, we assume~\cite{CGM78,YMA05} that the saddle of the form
\be
\phi^{\star}_{ab}\le(\mathbf{x}; \theta\ri)=\frac{1}{g}F\le(\mathbf{x}\ri) v^{(\theta)}_{ab}\, 
\ee
dictates the value $A(\theta)$. Here $F$ is a spherically symmetric function that solves
\be
\nabla^2F=F-F^2\, ,
\ee 
obtained numerically through the pseudospectral method~\cite{SpectralMatlab,AALY15}, and $v^{(\theta)}_{ab}$ is the replicon component of the Parisi RSB ansatz~\cite{Parisi79,MPV87,CC05,Denef11}. Computing the action of the resulting saddle (see Appendix~\ref{Bresum3}) indicates that a solution exists if and only if $1<\tan\theta<\infty$, with
\be
A\le(\theta\ri)=\frac{c_d }{\cos^2{\theta}\le(\tan\theta-1\ri)}\, ,
\ee
where $c_d$ is a $d$-dependent positive constant. The series is thus Borel-summable within the wedge $1<g^{\mathrm{II}}/g^{\mathrm{I}}<\infty$, consistent with the mean-field consideration~\cite{FPR96} and the two-loop basin of attraction obtained above. This result thus validates our perturbative treatment of the strong-coupling regime within the basin of attraction.

Given the large-order behavior at hand, we further compute the critical properties of the fixed point by resumming the three-loop series, analytically continuing the Borel transform through the conformal mapping~\cite{footnote_improve} (see Appendix~\ref{Bresum3}). Comparing the two-loop and the resummation results upon inclusion of higher-order contributions (Fig.~\ref{NFPinfo}) confirms that the fixed point is robustly conserved for $d\lesssim5.05$. The critical exponents from the two schemes further qualitatively agree with one another.

\emph{Conclusion--} The nontrivial critical fixed point identified here governs both dAT and Gardner transitions in $d<d_{\rm u}$. An RSB transition for the underlying universality class is thus possible over a broader $d$ range than previously thought~\cite{PT12,AB15,AB16}. The RG flow diagrams (Fig.~\ref{RGflow}) and the large-order behavior, however, make it clear that not all microscopic models belong to the basin of the attraction of the critical fixed point. This realization offers a possible explanation for the absence of dAT criticality in the Edwards-Anderson model in $d=3$. The model may simply remain outside the basin of attraction, and thus be governed either by a discontinuous transition into the RSB phase or by the two-state droplet picture~\cite{McMillan84,FH86,HF87,FH88,BM87}. Enlarging the range of disordered spin systems used for studying RSB criticality would clarify this last point.

Our results further highlight various future research directions. First, they guide efforts in systematizing nonperturbative RG methods~\cite{Wetterich93} and controlling conformal bootstrap techniques for nonunitary theories~\cite{RRTV08,EPPRSV12,Gliozzi13,EPPRSV14}. Both approaches should  find the nontrivial critical fixed point when applied to the replica field theory.
Second, conflicting results have been obtained for the lower critical dimension, $d_{\rm l}$, from a heuristic interface argument~\cite{FPV94} and from a correlation-function argument~\cite{DK84}. The dimensional dependence of the infrared divergence associated with soft modes thus deserves further scrutiny.
Third, extending the current approach will enable the study of the RG trajectory between the critical point identified here and the multicritical fixed point found perturbatively for the spin-glass transition in absence of external magnetic field~\cite{HLC76}, whereat longitudinal and anomalous modes become massless concurrently with the replicons.

\begin{acknowledgments}
We thank Giulio Biroli, Gerald V.~Dunne, Atsushi Ikeda, Shamit Kachru, Jaehoon Lee, Michael A.~Moore, Giorgio Parisi, Stephen H.~Shenker, Mithat \"{U}nsal, and Pierfrancesco Urbani for stimulating discussions and suggestions.
This work was supported by a grant from the Simons Foundation (\#454937, Patrick Charbonneau).
Data relevant to this work have been archived and can be accessed at http://dx.doi.org/10.7924/G86Q1V5C.
\end{acknowledgments}

\clearpage

\appendix
\renewcommand{\thefigure}{S\arabic{figure}}
\setcounter{figure}{0}
\begin{widetext}

\section{Replica field theory formalism}
\label{Lagrangian}
The basic object in the replica field theory is the overlap field, $q_{ab}\le(\mathbf{x}\ri)$, which is symmetric, $q_{ab}=q_{ba}$, and has no diagonal degree of freedom, \textit{i.e.}, $q_{aa}=0$, for replica indices $a,b$ running from $1$ to $n$.
The overlap field naturally decomposes into $1$ longitudinal, $(n-1)$ anomalous, and $n(n-3)/2$ replicon modes, for a total of $n(n-1)/2$ modes~\cite{NishimoriBook,BR80,PTD02}.
For perturbative calculations, we work within the critical surface on which replicon modes become massless while longitudinal and anomalous modes generically stay massive, and seek infrared stable fixed points within that surface.
For convenience we introduce an orthonormal basis of replicon modes that satisfies
\be
\sum_{a,b=1}^{n} e^{i}_{ab} e^{j}_{ab}=\delta^{ij}\, ,
\ee
where each vector has $e^{i}_{aa}=0$, and the replicon conditions require that
\be
\sum_{b=1}^{n} e^{i}_{ab}=0\, 
\ee
for all $a=1,\ldots,n$ and $i=1,\ldots,n(n-3)/2$.
With this notation, the replicon field can be written as
\be
\phi_{ab}\le(\mathbf{x}\ri)=\sum_{i=1}^{\frac{n(n-3)}{2}}\phi_i\le(\mathbf{x}\ri) e^{i}_{ab}\, 
\ee
and the bare massless Lagrangian as
\be
\mathcal{L}=\mathcal{L}_{\rm free}+\mathcal{L}_{\rm int}\, ,
\ee
where
\be
\mathcal{L}_{\rm free}=\frac{1}{2}\sum_{a,b=1}^{n}\le(\nabla\phi_{ab}\ri)^2=\frac{1}{2}\sum_{i=1}^{\frac{n(n-3)}{2}}\le(\nabla\phi_i\ri)^2
\ee
and
\bea
\mathcal{L}_{\rm int}&=&-\frac{1}{3!}\le(\gbare^{\rm I}\sum_{a,b=1}^n\phi_{ab}^3+\gbare^{\rm II}\sum_{a,b,c=1}^n\phi_{ab}\phi_{bc}\phi_{ca}\ri)\nonumber\\
&=&-\frac{1}{3!}\sum_{i,j,k=1}^{\frac{n(n-3)}{2}}\le(\gbare^{\rm I}T_{\rm I}^{ijk}+\gbare^{\rm II}T_{\rm II}^{ijk}\ri)\phi_{i}\phi_{j}\phi_{k}
\eea
with
\be
T_{\rm I}^{ijk}\equiv \sum_{a,b=1}^{n}e^{i}_{ab}e^{j}_{ab}e^{k}_{ab}\ \ \ \mathrm{and}\ \ \ T_{\rm II}^{ijk}\equiv \sum_{a,b,c=1}^{n}e^{i}_{ab}e^{j}_{bc}e^{k}_{ca}\, .
\ee
It has been shown in Ref.~\cite{PTD02} that, for replicon modes, these two terms exhaust all the cubic terms that are symmetric under the permutation of replica indices. Note that: (i) the action is symmetric under inverting the couplings $\le(\gbare^{\mathrm{I}},\gbare^{\mathrm{II}}\ri)\rightarrow \le(-\gbare^{\mathrm{I}},-\gbare^{\mathrm{II}}\ri)$ combined with $\phi_{ab}\rightarrow-\phi_{ab}$; (ii) a linear term is absent due to the replicon condition; (iii) the relevant quadratic mass term is suppressed for the calculations of $\beta$-functions within the critical surface but later added to calculate the critical exponent $\nu$ in Sec.~\ref{CE}; (iv) unlike in the Ising model, the cubic terms are \textit{not} redundant even in the presence of higher-order terms, because one cannot shift replicon modes by a constant; and (v) the cubic field theory has an upper critical dimension $d_{\rm u}=6$ and the Gaussian fixed point becomes unstable for $\epsilon\equiv d_{\rm u}-d>0$. Although cubic field theories with different symmetry structures have been studied up to four-loop order~\cite{Amit76,BKM81,Gracey15}, these results are here of limited interest because their actions belong to different universality classes.

The zero replica limit, $n\rightarrow 0$, is taken at the end of the calculation in order to properly average over disorder.
For spin glasses, disorder comes from quenched impurities. For structural glasses, by contrast, disorder is self-induced by reference configurations. For instance, one can follow the overlap between a reference configuration sampled right before falling out of equilibrium and replicated configurations out of equilibrium at a lower temperature or higher density, as well as the overlap among the latter set of configurations.
This scheme results in the state-following ensemble, in which the overlap field is an $(m+n)$-by-$(m+n)$ matrix field with $n\rightarrow0$ and $m=1$. Because replicon modes responsible for the Gardner transition reside within the $n$-by-$n$ submatrix, we can set the relevant number of replicas to $0$ for calculations within this ensemble. For the so-called Edwards-Monasson ensemble, however, the number of replicas at the Gardner transition can lie anywhere within the range $[0,1]$. Given that sampling protocols and experimental relevance of the Edwards-Monasson ensemble are less transparent than for the state-following ensemble, we here present results for the zero replica limit $n=0$ only. Note, however, that we do not observe any qualitative changes as a function of the number of replicas, except when $n$ is near $1$. The topology of fixed points then becomes complex, at least within the two-loop renormalization group scheme. In some dimensions $d$, the nontrivial fixed point disappears; in other dimensions, additional fixed points appear. In particular, just below $d_{\rm u}=6$, the finely-tuned fixed point first found in Ref.~\cite{BU15} emerges near $n\approx 0.90$. Although of certain interest, this fixed point appears to be nongeneric.

\section{Two-loop $\beta$-functions}
\label{2beta}

In this section we derive two-loop $\beta$-functions for the replica field theory within the dimensional regularization scheme.
The results of computations for two-loop amplitudes are reported in Sec.~\ref{Feynman}.
Dimensional regularization is implemented in Sec.~\ref{Regularization}, culminating with the derivation of the $\beta$-functions in Sec.~\ref{beta}.
Sec.~\ref{realhorror} lists various replica combinatorial factors.

\subsection{Feynman diagrams}
\label{Feynman}
To derive $\beta$-functions within the dimensional regularization scheme, we only need to calculate a few class of amplitudes associated with one-particle irreducible Feynman diagrams.
Specifically, we need the bare self-energy,
\be
\Pi^{ij}_{\mathrm{B}}(\mathbf{k})=\mathbf{k}^2 \delta^{ij} \tilde{\Pi}_{\mathrm{B}}(\mathbf{k})\, ,
\ee
which is given by the sum of all one-particle-irreducible Feynman diagrams with two external legs (Fig.~\ref{FeynmanSelf}), and the bare cubic vertex given by the sum of all one-particle-irreducible Feynman diagrams with three external legs (Fig.~\ref{FeynmanCubic}),
\be
-\mu^{\frac{\epsilon}{2}}\Gamma^{(3)ijk}_{\mathrm{B}}\le(\mathbf{k}_1,\mathbf{k}_2\ri)=-\mu^{\frac{\epsilon}{2}}\le\{\Gamma^{\rm I}_{\mathrm{B}}\le(\mathbf{k}_1,\mathbf{k}_2\ri)T_{\rm I}^{ijk}+\Gamma^{\rm II}_{\mathrm{B}}\le(\mathbf{k}_1,\mathbf{k}_2\ri)T_{\rm II}^{ijk}\ri\}\, .
\ee
We introduced the renormalization energy scale $\mu$, making $\Gamma_{\mathrm{B}}$'s dimensionless. To further lighten the notation, we introduce dimensionless bare couplings
\be\label{dimlessbare}
u_{\rm B}^{\T}\equiv \mu^{-\frac{\epsilon}{2}}\gbare^{\T}\, 
\ee
for $\T\in\le\{\mathrm{I}, \mathrm{II}\ri\}$.
The above amplitudes can formally be expanded in perturbative series as
\be
\tilde{\Pi}_{\mathrm{B}}=\sum_{m=0}^2 h^Z_{2-m, m} \le(u_{\rm B}^{\rm I}\ri)^{2-m}\le(u_{\rm B}^{\rm II}\ri)^{m}+\sum_{m=0}^4 h^Z_{4-m, m}\le(u_{\rm B}^{\rm I}\ri)^{4-m}\le(u_{\rm B}^{\rm II}\ri)^{m}+O(u_{\rm B}^6)\, 
\ee
and
\be
\Gamma^{\T}_{\mathrm{B}}=u_{\rm B}^{\T}+\sum_{m=0}^3 h^{\T}_{3-m, m} \le(u_{\rm B}^{\rm I}\ri)^{3-m}\le(u_{\rm B}^{\rm II}\ri)^{m}+\sum_{m=0}^5 h^{\T}_{5-m, m} \le(u_{\rm B}^{\rm I}\ri)^{5-m}\le(u_{\rm B}^{\rm II}\ri)^{m}+O(u_{\rm B}^7)\, .
\ee
The series coefficients, $\le\{h^Z_{m_1,m_2}\ri\}$ and $\le\{h^{\T}_{m_1,m_2}\ri\}$, are what need to be calculated.

Besides the tedious replica combinatorial factors and the presence of two distinct cubic couplings, factors arising from momentum integrals are the same as for simpler cubic theories, as worked out in Ref.~\cite{Amit76} to two-loop order, in Ref.~\cite{BKM81} to three-loop order, and in Ref.~\cite{Gracey15} to four-loop order.
Defining the common factor arising from the spherical integral, \textit{i.e.} the volume of a unit sphere divided by the Fourier factor for a loop-momentum integral, 
\be\label{sphere}
K_{d}\equiv\frac{\mathrm{vol}(S^{d-1})}{(2\pi)^d}=\frac{1}{2^{d-1}\pi^{\frac{d}{2}}\Gamma\le(\frac{d}{2}\ri)}\, ,
\ee
and momentum-dependent functions
\be
 L_0\equiv\mathrm{ln}\le(\frac{\mathbf{k}^2}{\mu^2}\ri)\, 
\ee
and
\be
L_1\equiv \int_0^1 \mathrm{d}x \int_0^{1-x}\mathrm{d}y\ \mathrm{ln}\le\{x(1-x)\frac{\mathbf{k}_1^2}{\mu^2}+y(1-y)\frac{\mathbf{k}_2^2}{\mu^2}+2xy\frac{\mathbf{k}_1\cdotp\mathbf{k}_2}{\mu^2} \ri\}    \, ,
\ee
amplitudes that appear in the cubic theories up to two-loop order are~\cite{Amit76}
\bea
A^{(2)}_1&\equiv&\frac{\mu^{\epsilon}}{\mathbf{k}^2}\int\frac{\mathrm{d}\mathbf{q}}{(2\pi)^d}\frac{1}{\mathbf{q}^2(\mathbf{q}+\mathbf{k})^2}\\
&=&K_{d}\le(\frac{-1}{3\epsilon}\ri)\le(1+\frac{7}{12}\epsilon-\frac{1}{2}\epsilon L_0\ri)+O\le(\epsilon\ri)\, ,\nonumber\\
A^{(3)}_1&\equiv&\mu^{\epsilon}\int\frac{\mathrm{d}\mathbf{q}}{(2\pi)^d}\frac{1}{\mathbf{q}^2(\mathbf{q}+\mathbf{k}_1)^2(\mathbf{q}+\mathbf{k}_1+\mathbf{k}_2)^2}\\
&=&K_{d}\le(\frac{1}{\epsilon}\ri)\le(1-\frac{3}{4}\epsilon-\epsilon L_1\ri)+O\le(\epsilon\ri)\, ,\nonumber\\
A^{(2)}_2&\equiv&\frac{\mu^{2\epsilon}}{\mathbf{k}^2}\int\frac{\mathrm{d}\mathbf{q}_1}{(2\pi)^d}\int\frac{\mathrm{d}\mathbf{q}_2}{(2\pi)^d}\frac{1}{\mathbf{q}_1^4(\mathbf{q}_1+\mathbf{k})^2\mathbf{q}_2^2(\mathbf{q}_1+\mathbf{q}_2)^2}\\
&=&K^2_{d}\le(\frac{1}{18\epsilon^2}\ri)\le(1+\frac{25}{12}\epsilon-\epsilon L_0\ri)+O\le(1\ri)\, ,\nonumber\\
B^{(2)}_2&\equiv&\frac{\mu^{2\epsilon}}{\mathbf{k}^2}\int\frac{\mathrm{d}\mathbf{q}_1}{(2\pi)^d}\int\frac{\mathrm{d}\mathbf{q}_2}{(2\pi)^d}\frac{1}{\mathbf{q}_1^2(\mathbf{q}_1+\mathbf{k})^2\mathbf{q}_2^2(\mathbf{q}_1+\mathbf{q}_2)^2(\mathbf{q}_1+\mathbf{q}_2+\mathbf{k})^2}\\
&=&K^2_{d}\le(\frac{-1}{3\epsilon^2}\ri)\le(1+\frac{3}{2}\epsilon-\epsilon L_0\ri)+O\le(1\ri)\, ,\nonumber\\
A^{(3)}_2&\equiv&\mu^{2\epsilon}\int\frac{\mathrm{d}\mathbf{q}_1}{(2\pi)^d}\int\frac{\mathrm{d}\mathbf{q}_2}{(2\pi)^d}\frac{1}{\mathbf{q}_1^4(\mathbf{q}_1+\mathbf{k}_1)^2(\mathbf{q}_1+\mathbf{k}_1+\mathbf{k}_2)^2\mathbf{q}_2^2(\mathbf{q}_1+\mathbf{q}_2)^2}\\
&=&K^2_{d}\le(\frac{-1}{6\epsilon^2}\ri)\le(1-\frac{11}{12}\epsilon-2\epsilon L_1\ri)+O\le(1\ri)\, ,\nonumber\\
B^{(3)}_2&\equiv&\mu^{2\epsilon}\int\frac{\mathrm{d}\mathbf{q}_1}{(2\pi)^d}\int\frac{\mathrm{d}\mathbf{q}_2}{(2\pi)^d}\frac{1}{\mathbf{q}_1^2(\mathbf{q}_1+\mathbf{k}_1)^2(\mathbf{q}_1-\mathbf{q}_2)^2\mathbf{q}_2^2(\mathbf{q}_2+\mathbf{k}_1)^2(\mathbf{q}_2+\mathbf{k}_1+\mathbf{k}_2)^2}\\
&=&K^2_{d}\le(\frac{1}{2\epsilon^2}\ri)\le(1-\frac{5}{4}\epsilon-2\epsilon L_1\ri)+O\le(1\ri)\, ,\ \ \ {\rm and}\nonumber\\
C^{(3)}_2&\equiv&\mu^{2\epsilon}\int\frac{\mathrm{d}\mathbf{q}_1}{(2\pi)^d}\int\frac{\mathrm{d}\mathbf{q}_2}{(2\pi)^d}\frac{1}{\mathbf{q}_1^2(\mathbf{q}_1+\mathbf{k}_1)^2(\mathbf{q}_1-\mathbf{q}_2)^2(\mathbf{q}_1-\mathbf{q}_2-\mathbf{k}_2)^2\mathbf{q}_2^2(\mathbf{q}_2+\mathbf{k}_1+\mathbf{k}_2)^2}\\
&=&K^2_{d}\le(\frac{1}{2\epsilon^2}\ri)\le(\epsilon\ri)+O\le(1\ri)\, .\nonumber
\eea

\floatsetup[figure]{style=plain,subcapbesideposition=top}
\begin{figure*}[t]
\centering{
\sidesubfloat[]{
\begin{fmffile}{A21}
\begin{fmfchar}(30,30)
\fmfleft{i1}
\fmfright{i2}
\fmf{plain,tension=5}{i1,v1}
\fmf{plain,tension=5}{i2,v2}
\fmf{plain,left}{v1,v2}
\fmf{plain,right}{v1,v2}
\end{fmfchar}
\end{fmffile}
}\quad%
\hspace{-0.3in}
\sidesubfloat[]{
\begin{fmffile}{A22}
\begin{fmfchar}(30,30)
\fmfleft{i1}
\fmfright{i2}
\fmf{plain,tension=5}{i1,v1}
\fmf{plain,tension=5}{i2,v2}
\fmf{plain,left}{v1,v2}
\fmf{plain}{v1,v3}
\fmf{plain}{v4,v2}
\fmf{plain,left,tension=0.5}{v3,v4}
\fmf{plain,right,tension=0.5}{v3,v4}
\end{fmfchar}
\end{fmffile}
}\quad%
\hspace{-0.3in}
\sidesubfloat[]{
\begin{fmffile}{B22}
\begin{fmfchar}(30,30)
\fmfleft{i1}
\fmfright{i2}
\fmftop{i3}
\fmfbottom{i4}
\fmf{plain,tension=5}{i1,v1}
\fmf{plain,tension=5}{i2,v2}
\fmf{plain}{v1,v3,v2}
\fmf{plain}{v1,v4,v2}
\fmf{plain,tension=0}{v3,v4}
\fmf{phantom}{v3,i3}
\fmf{phantom}{v4,i4}
\end{fmfchar}
\end{fmffile}
}
}
\vspace{-0.2in}
\caption{One-particle-irreducible Feynman diagrams with two external legs at one-loop (a) and two-loop [(b) and (c)] order.}
\label{FeynmanSelf}
\end{figure*}

\begin{figure*}[t]
\centering{
\sidesubfloat[]{
\begin{fmffile}{A31}
\begin{fmfchar}(30,40)
\fmftop{i1}
\fmfleft{i2}
\fmfright{i3}
\fmf{plain,tension=5}{i1,v1}
\fmf{plain,tension=5}{i2,v2}
\fmf{plain,tension=5}{i3,v3}
\fmf{plain}{v1,v2}
\fmf{plain,tension=1}{v2,v3}
\fmf{plain}{v3,v1}
\end{fmfchar}
\end{fmffile}
}\quad%
\hspace{-0.3in}
\sidesubfloat[]{
\begin{fmffile}{A32}
\begin{fmfchar}(30,40)
\fmftop{i1}
\fmfleft{i2}
\fmfright{i3}
\fmf{plain,tension=5}{i1,v1}
\fmf{plain,tension=5}{i2,v2}
\fmf{plain,tension=5}{i3,v3}
\fmf{plain}{v1,v2}
\fmf{plain,tension=2}{v2,v4}
\fmf{plain,left,tension=0.8}{v4,v5}
\fmf{plain,right,tension=0.8}{v4,v5}
\fmf{plain,tension=2}{v5,v3}
\fmf{plain}{v3,v1}
\end{fmfchar}
\end{fmffile}
}\quad%
\hspace{-0.3in}
\sidesubfloat[]{
\begin{fmffile}{B32}
\begin{fmfchar}(30,40)
\fmftop{i1}
\fmfleft{i2}
\fmfright{i3}
\fmf{plain,tension=5}{i1,v1}
\fmf{plain,tension=5}{i2,v2}
\fmf{plain,tension=5}{i3,v3}
\fmf{plain}{v1,v4,v2}
\fmf{plain}{v1,v5,v3}
\fmf{plain,tension=1}{v2,v3}
\fmf{plain,tension=0}{v4,v5}
\end{fmfchar}
\end{fmffile}
}\quad%
\hspace{-0.3in}
\sidesubfloat[]{
\begin{fmffile}{C32}
\begin{fmfchar}(30,40)
\fmftop{i1}
\fmfleft{i2}
\fmfright{i3}
\fmf{plain,tension=5}{i1,v1}
\fmf{plain,tension=3}{i2,v2}
\fmf{plain,tension=3}{i3,v3}
\fmf{plain,tension=2}{v1,v4}
\fmf{plain,tension=2}{v1,v5}
\fmf{plain,tension=0,rubout=5}{v4,v3}
\fmf{plain,tension=0}{v5,v2}
\fmf{plain}{v3,v5}
\fmf{plain}{v2,v4}
\fmf{phantom}{v4,i2}
\fmf{phantom}{v5,i3}
\end{fmfchar}
\end{fmffile}
}
}
\vspace{-0.7in}
\caption{One-particle-irreducible Feynman diagrams with three external legs at one-loop (a) and two-loop [(b)--(d)] order.}
\label{FeynmanCubic}
\end{figure*}
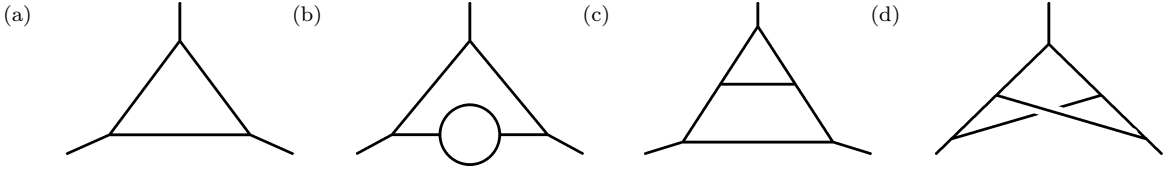

We then incorporate the replica combinatorial factors arising from the tensorial contractions of indices running in the loops, in addition to the standard symmetric factors associated with the respective Feynman diagrams.
The resulting one-loop self-energy [Fig.~\ref{FeynmanSelf}(a)] is
\be
\sum_{m=0}^2 h^Z_{2-m, m} \le(u_{\rm B}^{\rm I}\ri)^{2-m}\le(u_{\rm B}^{\rm II}\ri)^{m}=\frac{1}{2}A^{(2)}_1\le(\sum_{\T_1,\T_2\in\le\{\mathrm{I},\mathrm{II}\ri\}}S_{\T_1,\T_2}u_{\rm B}^{\T_1}u_{\rm B}^{\T_2}\ri)\, ,
\ee
where the replica combinatorial factors are defined through
\be
\label{defineS2}
\sum_{i_3,i_4=1}^{\frac{n(n-3)}{2}}T_{\T_1}^{i_1 i_3 i_4}T_{\T_2}^{i_2 i_4 i_3}\equiv S_{\T_1, \T_2} \delta^{i_1i_2}= S_{\T_2, \T_1} \delta^{i_1i_2}\, 
\ee
and their explicit expressions as functions of $n$ are listed in Eq.~\eqref{Ss}.
The one-loop cubic vertices [Fig.~\ref{FeynmanCubic}(a)] give
\be
\sum_{m=0}^3 h^{\T}_{3-m, m} \le(u_{\rm B}^{\rm I}\ri)^{3-m}\le(u_{\rm B}^{\rm II}\ri)^{m}=A^{(3)}_1\le(\sum_{\T_1,\T_2,\T_3\in\le\{\mathrm{I},\mathrm{II}\ri\}}a^{\T}_{\T_1,\T_2,\T_3}u_{\rm B}^{\T_1}u_{\rm B}^{\T_2}u_{\rm B}^{\T_3}\ri)\, ,
\ee
where replica combinatorial factors are defined through
\be
\label{definea3}
\sum_{i_4,i_5,i_6=1}^{\frac{n(n-3)}{2}}T_{\T_1}^{i_1 i_5 i_6}T_{\T_2}^{i_2 i_6 i_4}T_{\T_3}^{i_3 i_4 i_5} \equiv\sum_{\T\in\le\{\mathrm{I},\mathrm{II}\ri\}}a^{\T}_{\T_1,\T_2,\T_3}T_{\T}^{i_1 i_2 i_3}\, 
\ee
and symmetric under permutations of indices $(\T_1,\T_2,\T_3)$.
Their expressions are listed in Eq.~\eqref{a3s}. For later development, it is important to note that these coefficients satisfy the 't Hooft identities
\be
\sum_{\T\in\le\{\mathrm{I},\mathrm{II}\ri\}}S_{\T_1,\T}a^{\T}_{\T_2,\T_3,\T_4}=\sum_{\T\in\le\{\mathrm{I},\mathrm{II}\ri\}}S_{\T_2,\T}a^{\T}_{\T_1,\T_3,\T_4}\, ,
\ee
which can be derived diagrammatically by cutting appropriate two-loop self-energy diagrams in different ways and can also be checked explicitly through Eqs.~\eqref{Ss} and \eqref{a3s}.
With these identities in mind, the two-loop self-energy [Figs.~\ref{FeynmanSelf}(b) and (c)] gives
\bea
\sum_{m=0}^4 h^Z_{4-m, m}\le(u_{\rm B}^{\rm I}\ri)^{4-m}\le(u_{\rm B}^{\rm II}\ri)^{m}&=&\frac{1}{2}A^{(2)}_2 \le(\sum_{\T_1,\T_2\in\le\{\mathrm{I},\mathrm{II}\ri\}}S_{\T_1,\T_2}u_{\rm B}^{\T_1}u_{\rm B}^{\T_2}\ri)^2\\
&&+\frac{1}{2}B^{(2)}_2 \le(\sum_{\T_1,\T_2,\T_3,\T_4,\T_5\in\le\{\mathrm{I},\mathrm{II}\ri\}}S_{\T_1,\T_5}a^{\T_5}_{\T_2,\T_3,\T_4}u_{\rm B}^{\T_1}u_{\rm B}^{\T_2}u_{\rm B}^{\T_3}u_{\rm B}^{\T_4}\ri)\, .\nonumber
\eea
Finally, the two-loop cubic vertices [Figs.~\ref{FeynmanCubic}(b), (c), and (d)] give
\bea
\sum_{m=0}^5 h^{\T}_{5-m, m} \le(u_{\rm B}^{\rm I}\ri)^{5-m}\le(u_{\rm B}^{\rm II}\ri)^{m}&=&
\frac{3}{2}A^{(3)}_2  \le(\sum_{\T_1,\T_2\in\le\{\mathrm{I},\mathrm{II}\ri\}}S_{\T_1,\T_2}u_{\rm B}^{\T_1}u_{\rm B}^{\T_2}\ri)\le(\sum_{\T_3,\T_4,\T_5\in\le\{\mathrm{I},\mathrm{II}\ri\}}a^{\T}_{\T_3,\T_4,\T_5}u_{\rm B}^{\T_3}u_{\rm B}^{\T_4}u_{\rm B}^{\T_5}\ri)\\
  &&+3B^{(3)}_2\le(\sum_{\T_1,\T_2,\T_3,\T_4,\T_5,\T_6\in\le\{\mathrm{I},\mathrm{II}\ri\}}a^{\T}_{\T_1,\T_2,\T_6}a^{\T_6}_{\T_3,\T_4,\T_5}u_{\rm B}^{\T_1}u_{\rm B}^{\T_2}u_{\rm B}^{\T_3}u_{\rm B}^{\T_4}u_{\rm B}^{\T_5}\ri)\nonumber\\
  &&+\frac{1}{2}C^{(3)}_2\le(\sum_{\T_1,\T_2,\T_3,\T_4,\T_5\in\le\{\mathrm{I},\mathrm{II}\ri\}}a^{\T}_{\T_1,\T_2,\T_3;\T_4,\T_5}u_{\rm B}^{\T_1}u_{\rm B}^{\T_2}u_{\rm B}^{\T_3}u_{\rm B}^{\T_4}u_{\rm B}^{\T_5}\ri)\, , \nonumber
\eea
where we introduced new replica combinatorial factors of the form
\be
\label{definea5}
\sum_{i_4,i_5,i_6,i_7,i_8,i_9=1}^{\frac{n(n-3)}{2}}T_{\T_1}^{i_1 i_5 i_6}T_{\T_2}^{i_2 i_4 i_8}T_{\T_3}^{i_3 i_7 i_9}T_{\T_4}^{i_4 i_6 i_9}T_{\T_5}^{i_5 i_7 i_8}
\equiv \sum_{\T\in\le\{\mathrm{I},\mathrm{II}\ri\}}a^{\T}_{\T_1,\T_2,\T_3;\T_4,\T_5} T_{\T}^{i_1 i_2 i_3}\, 
\ee
with values listed in Eqs.~\eqref{a5sI} and \eqref{a5sII}. Note that they are symmetric under permutations of the first three indices $(\T_1,\T_2,\T_3)$ and of the last two indices $(\T_4,\T_5)$.

\subsection{Dimensional regularization}
\label{Regularization}
All the bare amplitudes, $\tilde{\Pi}_{\mathrm{B}}$ and $\Gamma^{\T}_{\mathrm{B}}$, diverge as $\epsilon\rightarrow0^{+}$ (with the ultraviolet cutoff for the momentum integral, $\Lambda$, kept infinite), but dimensional regularization tames these infinities~\cite{HV72,H73a,H73b}.
In this scheme, bare couplings are first expanded in terms of physical running couplings, $\le\{g_{\rm P}^{\mathrm{I}},g_{\rm P}^{\mathrm{II}}\ri\}$, as
\be\label{PBrelation}
K_d^{\frac{1}{2}}u_{\rm B}^{\T}=g_{\rm P}^{\T}+\sum_{m=0}^3 f^{\T}_{3-m, m} \le(g_{\rm P}^{\rm I}\ri)^{3-m}\le(g_{\rm P}^{\rm II}\ri)^{m}+\sum_{m=0}^5 f^{\T}_{5-m, m} \le(g_{\rm P}^{\rm I}\ri)^{5-m}\le(g_{\rm P}^{\rm II}\ri)^{m}+O(g_{\rm P}^7)\, ,
\ee
and the field is renormalized by introducing the wavefunction renormalization factor,
\be
Z_{\phi}=1+\sum_{m=0}^2 f^Z_{2-m, m} \le(g_{\rm P}^{\rm I}\ri)^{2-m}\le(g_{\rm P}^{\rm II}\ri)^{m}+\sum_{m=0}^4 f^Z_{4-m, m} \le(g_{\rm P}^{\rm I}\ri)^{4-m}\le(g_{\rm P}^{\rm II}\ri)^{m}+O(g_{\rm P}^6)\, .
\ee
Note that we have factored out the spherical factor $K_d^{\frac{1}{2}}$ from cubic couplings, as is conventionally done.
We then regulate divergences by adjusting series coefficients so as to keep the following renormalized physical amplitudes finite in the limit $\epsilon\rightarrow0$, \textit{i.e.},
\be\label{R1}
\Gamma^{(2)}_{\mathrm{P}}\le(\mathbf{k}\ri)\equiv Z_{\phi}\le\{1-\tilde{\Pi}_{\mathrm{B}}(\mathbf{k})\ri\}=\mathrm{finite}\, 
\ee
and
\be\label{R2}
\Gamma^{\T}_{\mathrm{P}}\le(\mathbf{k}_1,\mathbf{k}_2\ri)\equiv Z_{\phi}^{3/2}\le\{\Gamma^{\T}_{\mathrm{B}}\le(\mathbf{k}_1,\mathbf{k}_2\ri)\ri\}=\mathrm{finite}\, .
\ee
Specifically, order by order, we adjust the series coefficients, $\le\{f^Z_{m_1,m_2}\ri\}$ and $\le\{f^{\T}_{m_1,m_2}\ri\}$, so as to keep quantities \eqref{R1} and \eqref{R2} finite by \textit{minimally subtracting} the poles in $\epsilon$ stemming from those residing in $\le\{h^Z_{m_1,m_2}\ri\}$ and $\le\{h^{\T}_{m_1,m_2}\ri\}$.
This procedure renders all the other renormalized amplitudes of elementary operators finite (see Sec.~\ref{CE} for the case with a composite operator). 
We denote by $[\![\ldots]\!]_{\rm s}$  the singular terms proportional to poles in $\epsilon$ of the form $\epsilon^{-p}$ with $p>0$.

Implementing this scheme, the wavefunction renormalization condition of Eq.~\eqref{R1} at leading one-loop order yields
\bea
\sum_{m=0}^2 f^Z_{2-m, m} \le(g_{\rm P}^{\rm I}\ri)^{2-m}\le(g_{\rm P}^{\rm II}\ri)^{m}&=&\sum_{m=0}^2 [\![K_d^{-1}h^Z_{2-m, m}]\!]_{\rm s} \le(g_{\rm P}^{\rm I}\ri)^{2-m}\le(g_{\rm P}^{\rm II}\ri)^{m}\\
&=&\le(\frac{-1}{6\epsilon}\ri)\le(\sum_{\T_1,\T_2\in\le\{\mathrm{I},\mathrm{II}\ri\}}S_{\T_1,\T_2}g_{\rm P}^{\T_1}g_{\rm P}^{\T_2}\ri)\, .\nonumber
\eea
Similarly, the cubic vertex renormalization condition \eqref{R2} at one-loop order yield
\bea
\sum_{m=0}^3 f^{\T}_{3-m, m} \le(g_{\rm P}^{\rm I}\ri)^{3-m}\le(g_{\rm P}^{\rm II}\ri)^{m}&=&
-\sum_{m=0}^3 [\![K_d^{-1}h^{\T}_{3-m, m}]\!]_{\rm s} \le(g_{\rm P}^{\rm I}\ri)^{3-m}\le(g_{\rm P}^{\rm II}\ri)^{m}-\frac{3}{2}g_{\rm P}^{\T}\sum_{m=0}^2 [\![f^Z_{2-m, m}]\!]_{\rm s} \le(g_{\rm P}^{\rm I}\ri)^{2-m}\le(g_{\rm P}^{\rm II}\ri)^{m}\\
&=&\le(\frac{-1}{\epsilon}\ri)\le(\sum_{\T_1,\T_2,\T_3\in\le\{\mathrm{I},\mathrm{II}\ri\}}a^{\T}_{\T_1,\T_2,\T_3}g_{\rm P}^{\T_1}g_{\rm P}^{\T_2}g_{\rm P}^{\T_3}\ri)
+\le(\frac{1}{4\epsilon}\ri)g_{\rm P}^{\T}\le(\sum_{\T_1,\T_2\in\le\{\mathrm{I},\mathrm{II}\ri\}}S_{\T_1,\T_2}g_{\rm P}^{\T_1}g_{\rm P}^{\T_2}\ri)\, .\nonumber
\eea
Continuing to subleading two-loop order, the wavefunction renormalization condition \eqref{R1} yields
\bea
\sum_{m=0}^4 f^Z_{4-m, m} \le(g_{\rm P}^{\rm I}\ri)^{4-m}\le(g_{\rm P}^{\rm II}\ri)^{m}&=&\le(\frac{-1}{36\epsilon^2}\ri)\le(1-\frac{11}{12}\epsilon\ri)\le(\sum_{\T_1,\T_2\in\le\{\mathrm{I},\mathrm{II}\ri\}}S_{\T_1,\T_2}g_{\rm P}^{\T_1}g_{\rm P}^{\T_2}\ri)^2\\
&&+\le(\frac{1}{6\epsilon^2}\ri)\le(1-\frac{1}{3}\epsilon\ri) \le(\sum_{\T_1,\T_2,\T_3,\T_4,\T_5\in\le\{\mathrm{I},\mathrm{II}\ri\}}S_{\T_1,\T_5}a^{\T_5}_{\T_2,\T_3,\T_4}g_{\rm P}^{\T_1}g_{\rm P}^{\T_2}g_{\rm P}^{\T_3}g_{\rm P}^{\T_4}\ri)\, .\nonumber
\eea
Finally, Eq.~\eqref{R2} at two-loop order yields
\bea
\sum_{m=0}^5 f^{\T}_{5-m, m} \le(g_{\rm P}^{\rm I}\ri)^{5-m}\le(g_{\rm P}^{\rm II}\ri)^{m}&=&
\le(\frac{1}{16\epsilon^2}\ri)\le(\frac{3}{2}-\frac{11}{18}\epsilon\ri)g_{\rm P}^{\T}\le(\sum_{\T_1,\T_2\in\le\{\mathrm{I},\mathrm{II}\ri\}}S_{\T_1,\T_2}g_{\rm P}^{\T_1}g_{\rm P}^{\T_2}\ri)^2\\
&&+\le(\frac{-1}{4\epsilon^2}\ri)\le(1-\frac{1}{3}\epsilon\ri)g_{\rm P}^{\T} \le(\sum_{\T_1,\T_2,\T_3,\T_4,\T_5\in\le\{\mathrm{I},\mathrm{II}\ri\}}S_{\T_1,\T_5}a^{\T_5}_{\T_2,\T_3,\T_4}g_{\rm P}^{\T_1}g_{\rm P}^{\T_2}g_{\rm P}^{\T_3}g_{\rm P}^{\T_4}\ri)\nonumber\\
&&+\le(\frac{-1}{2\epsilon^2}\ri)\le(1-\frac{7}{24}\epsilon\ri)\le(\sum_{\T_1,\T_2,\T_3\in\le\{\mathrm{I},\mathrm{II}\ri\}}a^{\T}_{\T_1,\T_2,\T_3}g_{\rm P}^{\T_1}g_{\rm P}^{\T_2}g_{\rm P}^{\T_3}\ri)\le(\sum_{\T_4,\T_5\in\le\{\mathrm{I},\mathrm{II}\ri\}}S_{\T_4,\T_5}g_{\rm P}^{\T_4}g_{\rm P}^{\T_5}\ri)\nonumber\\
&&+\le(\frac{3}{2\epsilon^2}\ri)\le(1-\frac{1}{4}\epsilon\ri)\le(\sum_{\T_1,\T_2,\T_3,\T_4,\T_5,\T_6\in\le\{\mathrm{I},\mathrm{II}\ri\}}a^{\T}_{\T_1,\T_2,\T_6}a^{\T_6}_{\T_3,\T_4,\T_5}g_{\rm P}^{\T_1}g_{\rm P}^{\T_2}g_{\rm P}^{\T_3}g_{\rm P}^{\T_4}g_{\rm P}^{\T_5}\ri)\nonumber\\
&&+\le(\frac{-1}{4\epsilon}\ri)\le(\sum_{\T_1,\T_2,\T_3,\T_4,\T_5\in\le\{\mathrm{I},\mathrm{II}\ri\}}a^{\T}_{\T_1,\T_2,\T_3;\T_4,\T_5}g_{\rm P}^{\T_1}g_{\rm P}^{\T_2}g_{\rm P}^{\T_3}g_{\rm P}^{\T_4}g_{\rm P}^{\T_5}\ri)\, .\nonumber
\eea
We note that series coefficients, $\le\{f^Z_{m_1,m_2}\ri\}$ and $\le\{f^{\T}_{m_1,m_2}\ri\}$, are independent of momenta due to the cancellation of all momentum-dependent terms. This cancellation provides independent and highly nontrivial checks of the algebra and is one of many practical advantages of dimensional regularization.

\subsection{$\beta$-functions}
\label{beta}
The previous sections provide the necessary ingredients for obtaining two-loop expressions for $\beta$-functions,
\be
\beta^{\T}\equiv\mu\frac{\partial g_{\rm P}^{\T}}{\partial \mu}=-\frac{\epsilon}{2}g_{\rm P}^{\T}+\sum_{m=0}^3 \beta^{\T}_{3-m, m} \le(g_{\rm P}^{\rm I}\ri)^{3-m}\le(g_{\rm P}^{\rm II}\ri)^{m}+\sum_{m=0}^5 \beta^{\T}_{5-m, m} \le(g_{\rm P}^{\rm I}\ri)^{5-m}\le(g_{\rm P}^{\rm II}\ri)^{m}+O(g_{\rm P}^7)\, .
\ee
In order to obtain coefficients $\beta^{\T}_{m_1,m_2}$, we (i) express $\le\{g^{\T}_{\rm P}\ri\}_{\T\in\le\{\mathrm{I},\mathrm{II}\ri\}}$ in terms of $\le\{u^{\T}_{\rm B}\ri\}_{\T\in\le\{\mathrm{I},\mathrm{II}\ri\}}$, (ii) use the identity
\be
\mu\frac{\partial \le\{\le(u_{\rm B}^{\rm I}\ri)^{m_1}\le(u_{\rm B}^{\rm II}\ri)^{m_2}\ri\} }{\partial \mu}=-\frac{(m_1+m_2)\epsilon}{2}\le(u_{\rm B}^{\rm I}\ri)^{m_1}\le(u_{\rm B}^{\rm II}\ri)^{m_2}
\ee
which follows from the requirement that microscopic couplings must be independent of probing energy scale, \textit{i.e.}, $\frac{\partial \gbare^{\T}}{\partial \mu}=0$, and (iii) re-express $\le\{u^{\T}_{\rm B}\ri\}_{\T\in\le\{\mathrm{I},\mathrm{II}\ri\}}$ in terms of $\le\{g^{\T}_{\rm P}\ri\}_{\T\in\le\{\mathrm{I},\mathrm{II}\ri\}}$.
Straightforward algebra then yields
\bea
\beta_{\mathrm{1-loop}}^{\T}&\equiv&\sum_{m=0}^3 \beta^{\T}_{3-m, m} \le(g_{\rm P}^{\rm I}\ri)^{3-m}\le(g_{\rm P}^{\rm II}\ri)^{m}\\
&=&\epsilon\sum_{m=0}^3 f^{\T}_{3-m, m} \le(g_{\rm P}^{\rm I}\ri)^{3-m}\le(g_{\rm P}^{\rm II}\ri)^{m}\nonumber\\
&=&\frac{1}{4}g_{\rm P}^{\T}\le(\sum_{\T_1,\T_2\in\le\{\mathrm{I},\mathrm{II}\ri\}}S_{\T_1,\T_2}g_{\rm P}^{\T_1}g_{\rm P}^{\T_2}\ri)-\le(\sum_{\T_1,\T_2,\T_3\in\le\{\mathrm{I},\mathrm{II}\ri\}}a^{\T}_{\T_1,\T_2,\T_3}g_{\rm P}^{\T_1}g_{\rm P}^{\T_2}g_{\rm P}^{\T_3}\ri)\, \nonumber
\eea
at one-loop order, and at two-loop order [see Eq.~\eqref{3loopbeta} for a more concise expression]%there were typoes in ver1, and $m_2$ should have run from $-1$ to $3$, NOT from $0$ to $3$.
\bea
\sum_{m=0}^5 \beta^{\T}_{5-m, m} \le(g_{\rm P}^{\rm I}\ri)^{5-m}\le(g_{\rm P}^{\rm II}\ri)^{m}&=&2\epsilon\sum_{m=0}^5 f^{\T}_{5-m, m} \le(g_{\rm P}^{\rm I}\ri)^{5-m}\le(g_{\rm P}^{\rm II}\ri)^{m}-\frac{1}{\epsilon}\sum_{\T_1\in\le\{\mathrm{I},\mathrm{II}\ri\}}\frac{\partial\beta_{\mathrm{1-loop}}^{\T}}{\partial g_{\rm P}^{\T_1}}\partial\beta_{\mathrm{1-loop}}^{\T_1}\\
&=&\le(\frac{-11}{144}\ri)g_{\rm P}^{\T}\le(\sum_{\T_1,\T_2\in\le\{\mathrm{I},\mathrm{II}\ri\}}S_{\T_1,\T_2}g_{\rm P}^{\T_1}g_{\rm P}^{\T_2}\ri)^2\nonumber\\
&&+\frac{1}{6}g_{\rm P}^{\T} \le(\sum_{\T_1,\T_2,\T_3,\T_4,\T_5\in\le\{\mathrm{I},\mathrm{II}\ri\}}S_{\T_1,\T_5}a^{\T_5}_{\T_2,\T_3,\T_4}g_{\rm P}^{\T_1}g_{\rm P}^{\T_2}g_{\rm P}^{\T_3}g_{\rm P}^{\T_4}\ri)\nonumber\\
&&+\frac{7}{24}\le(\sum_{\T_1,\T_2,\T_3\in\le\{\mathrm{I},\mathrm{II}\ri\}}a^{\T}_{\T_1,\T_2,\T_3}g_{\rm P}^{\T_1}g_{\rm P}^{\T_2}g_{\rm P}^{\T_3}\ri)\le(\sum_{\T_4,\T_5\in\le\{\mathrm{I},\mathrm{II}\ri\}}S_{\T_4,\T_5}g_{\rm P}^{\T_4}g_{\rm P}^{\T_5}\ri)\nonumber\\
&&+\le(\frac{-3}{4}\ri)\le(\sum_{\T_1,\T_2,\T_3,\T_4,\T_5,\T_6\in\le\{\mathrm{I},\mathrm{II}\ri\}}a^{\T}_{\T_1,\T_2,\T_6}a^{\T_6}_{\T_3,\T_4,\T_5}g_{\rm P}^{\T_1}g_{\rm P}^{\T_2}g_{\rm P}^{\T_3}g_{\rm P}^{\T_4}g_{\rm P}^{\T_5}\ri)\nonumber\\
&&+\le(\frac{-1}{2}\ri)\le(\sum_{\T_1,\T_2,\T_3,\T_4,\T_5\in\le\{\mathrm{I},\mathrm{II}\ri\}}a^{\T}_{\T_1,\T_2,\T_3;\T_4,\T_5}g_{\rm P}^{\T_1}g_{\rm P}^{\T_2}g_{\rm P}^{\T_3}g_{\rm P}^{\T_4}g_{\rm P}^{\T_5}\ri)\, .\nonumber
\eea
Note that the coefficients, $\le\{\beta^{\T}_{m_1,m_2}\ri\}$, are all independent of $\epsilon$ and thus the $\epsilon$-dependence only appears in the first term of each $\beta$-function. This cancellation provides another set of highly nontrivial checks for higher-loop calculations in the dimensional regularization scheme.

\subsection{Combinatorial factors}
\label{realhorror}
The various combinatorial factors were obtained by implementing the following numerical algorithm: (i) form an orthonormal basis of replicon modes, $\le\{e^{i}_{ab}\ri\}_{i=1,\ldots,n(n-3)/2}$, through the Gram-Schmidt process; (ii) evaluate cubic generators, $T_{\rm I}^{ijk}$ and $T_{\rm II}^{ijk}$; (iii) obtain self-energy combinatorial factors for various $n$ as in Eq.~\eqref{defineS2} by evaluating diagonal components; (iv) obtain cubic combinatorial factors for various $n$ as in Eqs.~\eqref{definea3} and \eqref{definea5} by evaluating the equation of interest for two distinct set of indices, $(i_1i_2i_3)=(i^{(1)}_1i^{(1)}_2i^{(1)}_3),(i^{(2)}_1i^{(2)}_2i^{(2)}_3)$, such that $\le(T_{\rm I}^{i^{(1)}_1i^{(1)}_2i^{(1)}_3},T_{\rm II}^{i^{(1)}_1i^{(1)}_2i^{(1)}_3}\ri)$ and $\le(T_{\rm I}^{i^{(2)}_1i^{(2)}_2i^{(2)}_3},T_{\rm II}^{i^{(2)}_1i^{(2)}_2i^{(2)}_3}\ri)$ are linearly-independent; (v) fit combinatorial factors obtained for various $n$ ($n=6,\ldots,15$ suffices for our purpose) by rational functions with integer coefficients, noting that the denominator has the form $2^p(n-1)^q (n-2)^r$; and (vi) validate the consistency of the expressions by repeating the calculations up to $n=22$. The results follow.
\be\label{Ss}
\left[ {\begin{array}{c}
   S_{{\rm I},{\rm I}}  \\
   S_{{\rm I},{\rm II}}  \\
   S_{{\rm II},{\rm II}}   \\
  \end{array} } \right]=\left[ {\begin{array}{c}
   \frac{n^3-9n^2+26n-22}{2(n-1)(n-2)^2}  \\
   \frac{3n^2-15n+16}{2(n-1)(n-2)^2}  \\
   \frac{n^4-8n^3+19n^2-4n-16}{4(n-1)(n-2)^2}   \\
  \end{array} } \right]\, .
\ee
\be\label{a3s}
\left[ {\begin{array}{cc}
   a^{\mathrm{I}}_{{\rm I},{\rm I},{\rm I}} & a^{\mathrm{II}}_{{\rm I},{\rm I},{\rm I}} \\
   a^{\mathrm{I}}_{{\rm I},{\rm I},{\rm II}} & a^{\mathrm{II}}_{{\rm I},{\rm I},{\rm II}} \\
  a^{\mathrm{I}}_{{\rm I},{\rm II},{\rm II}}  & a^{\mathrm{II}}_{{\rm I},{\rm II},{\rm II}} \\
  a^{\mathrm{I}}_{{\rm II},{\rm II},{\rm II}} & a^{\mathrm{II}}_{{\rm II},{\rm II},{\rm II}} \\
  \end{array} } \right]=\left[ {\begin{array}{cc}
    \frac{n^3-11n^2+38n-34}{2(n-1)(n-2)^2} & \frac{-1}{(n-2)^3} \\
    \frac{3n^2-19n+20}{2(n-1)(n-2)^2}  & \frac{-n^3+8n^2-17n+12}{2(n-1)(n-2)^3} \\
   \frac{-n^3+5n^2+8n-16}{4(n-1)(n-2)^2}  & \frac{3n^3-27n^2+64n-48}{4(n-1)(n-2)^3} \\
   \frac{-3n}{2(n-2)^2}  & \frac{n^5-10n^4+33n^3-8n^2-104n+112}{8(n-1)(n-2)^3} \\
  \end{array} } \right]\, .
\ee
\be
\label{a5sI}
\left[ {\begin{array}{c}
  a^{\mathrm{I}}_{{\rm I},{\rm I},{\rm I};{\rm I},{\rm I}}  \\
  a^{\mathrm{I}}_{{\rm II},{\rm I},{\rm I};{\rm I},{\rm I}} \\
  a^{\mathrm{I}}_{{\rm I},{\rm I},{\rm I};{\rm II},{\rm I}}  \\
  a^{\mathrm{I}}_{{\rm I},{\rm I},{\rm I};{\rm II},{\rm II}}  \\
  a^{\mathrm{I}}_{{\rm II},{\rm II},{\rm I};{\rm I},{\rm I}} \\
  a^{\mathrm{I}}_{{\rm II},{\rm I},{\rm I};{\rm II},{\rm I}}  \\
  a^{\mathrm{I}}_{{\rm II},{\rm II},{\rm II};{\rm I},{\rm I}}  \\
  a^{\mathrm{I}}_{{\rm I},{\rm I},{\rm II};{\rm II},{\rm II}}  \\
  a^{\mathrm{I}}_{{\rm I},{\rm II},{\rm II};{\rm I},{\rm II}} \\
  a^{\mathrm{I}}_{{\rm I},{\rm II},{\rm II};{\rm II},{\rm II}}  \\
  a^{\mathrm{I}}_{{\rm II},{\rm II},{\rm II};{\rm I},{\rm II}}  \\  
  a^{\mathrm{I}}_{{\rm II},{\rm II},{\rm II};{\rm II},{\rm II}} \\
  \end{array} } \right]
  \equiv
  \left[ {\begin{array}{c}
  \frac{n^8-26n^7+291n^6-1816n^5+6840n^4-15756n^3+21586n^2-16088n+5008}{4(n-1)^2 (n-2)^6}  \\
  \frac{3n^7-66n^6+607n^5-2960n^4+8132n^3-12592n^2+10236n-3392}{4(n-1)^2 (n-2)^6}  \\
  \frac{3n^7-66n^6+604n^5-2930n^4+8017n^3-12380n^2+10048n-3328}{4(n-1)^2 (n-2)^6}  \\
  \frac{21n^6-366n^5+2493n^4-8316n^3+14536n^2-12800n+4480}{8(n-1)^2 (n-2)^6}  \\
  \frac{3n^7-27n^6-59n^5+1471n^4-6396n^3+12496n^2-11664n+4224}{8(n-1)^2 (n-2)^6}  \\
  \frac{n^7-7n^6-63n^5+819n^4-3292n^3+6262n^2-5776n+2080}{4(n-1)^2 (n-2)^6}  \\
  \frac{n^9-19n^8+145n^7-541n^6+1018n^5-1488n^4+4292n^3-10192n^2+11328n-4608}{16(n-1)^2 (n-2)^6} \\
  \frac{-n^7+20n^6-110n^5+84n^4+871n^3-2704n^2+3040n-1216}{4(n-1)^2 (n-2)^6}  \\
  \frac{-7n^7+134n^6-819n^5+1708n^4+680n^3-7552n^2+10144n-4352}{16(n-1)^2 (n-2)^6}  \\
  \frac{n^9-15n^8+95n^7-469n^6+2196n^5-6368n^4+8592n^3-2176n^2-5376n+3584}{32(n-1)^2 (n-2)^6}  \\
  \frac{3n^8-42n^7+169n^6+68n^5-1750n^4+3488n^3-1456n^2-1984n+1536}{16(n-1)^2 (n-2)^6}  \\
  \frac{n(-3n^6+54n^5-315n^4+560n^3+376n^2-1968n+1440)}{16(n-1)(n-2)^6}  \\
    \end{array} } \right]\, .
\ee
\be
\label{a5sII}
\left[ {\begin{array}{c}
  a^{\mathrm{II}}_{{\rm I},{\rm I},{\rm I};{\rm I},{\rm I}}  \\
  a^{\mathrm{II}}_{{\rm II},{\rm I},{\rm I};{\rm I},{\rm I}} \\
  a^{\mathrm{II}}_{{\rm I},{\rm I},{\rm I};{\rm II},{\rm I}}  \\
  a^{\mathrm{II}}_{{\rm I},{\rm I},{\rm I};{\rm II},{\rm II}}  \\
  a^{\mathrm{II}}_{{\rm II},{\rm II},{\rm I};{\rm I},{\rm I}} \\
  a^{\mathrm{II}}_{{\rm II},{\rm I},{\rm I};{\rm II},{\rm I}}  \\
  a^{\mathrm{II}}_{{\rm II},{\rm II},{\rm II};{\rm I},{\rm I}}  \\
  a^{\mathrm{II}}_{{\rm I},{\rm I},{\rm II};{\rm II},{\rm II}}  \\
  a^{\mathrm{II}}_{{\rm I},{\rm II},{\rm II};{\rm I},{\rm II}} \\
  a^{\mathrm{II}}_{{\rm I},{\rm II},{\rm II};{\rm II},{\rm II}}  \\
  a^{\mathrm{II}}_{{\rm II},{\rm II},{\rm II};{\rm I},{\rm II}}  \\  
  a^{\mathrm{II}}_{{\rm II},{\rm II},{\rm II};{\rm II},{\rm II}} \\
  \end{array} } \right]
  \equiv
  \left[ {\begin{array}{c}
  \frac{3(n^2-7n+8)}{(n-1)^1 (n-2)^5} \\
  \frac{n^5-15n^4+78n^3-165n^2+159n-62}{2(n-1)^2 (n-2)^5} \\
   \frac{3n^5-42n^4+211n^3-448n^2+436n-168}{4(n-1)^2 (n-2)^5} \\
  \frac{n^7-18n^6+127n^5-420n^4+574n^3-40n^2-608n+416}{8(n-1)^2 (n-2)^5} \\
 \frac{-n^5+19n^4-118n^3+296n^2-336n+148}{2(n-1)^2 (n-2)^5} \\
   \frac{-2n^5+41n^4-260n^3+659n^2-750n+328}{4(n-1)^2 (n-2)^5} \\
    \frac{3n^5-72n^4+531n^3-1494n^2+1848n-864}{8(n-1)^2 (n-2)^5} \\
    \frac{3n^6-39n^5+151n^4-45n^3-726n^2+1344n-736}{8(n-1)^2 (n-2)^5} \\
  \frac{n^7-14n^6+81n^5-352n^4+1412n^3-3384n^2+3984n-1824}{16(n-1)^2 (n-2)^5} \\
 \frac{3n^5-17n^4-25n^3+243n^2-420n+232}{2(n-1)^2 (n-2)^5} \\
  \frac{3n^6-24n^5+147n^4-1006n^3+3136n^2-4240n+2112}{16(n-1)^2 (n-2)^5} \\
   \frac{3n^8-47n^7+315n^6-1229n^5+3110n^4-4088n^3+336n^2+4928n-3648}{32(n-1)^2(n-2)^5} \\
    \end{array} } \right]\, .
\ee

\section{Two-loop Critical exponents}
\label{CE}
In this section, we obtain two-loop expressions for the two critical exponents, $\eta$ and $\nu$.
The critical exponent $\eta$, which governs the decay of correlation functions right at the critical point, can be derived from the information obtained in Sec.~\ref{2beta} within the critical surface on which the mass of replicon modes stays strictly zero.
The critical exponent $\nu$, which governs the divergence of the correlation length as one approaches the critical surface, however, additionally requires the amplitudes with one insertion of the following \textit{composite} operator, corresponding to the relevant replicon-mass deformation
\be
\frac{1}{2}\sum_{a,b=1}^{n}\phi_{ab}^2=\frac{1}{2}\sum_{i=1}^{\frac{n(n-3)}{2}}\phi_i^2\, .
\ee
The bare amplitude is given by the sum of all the one-particle-irreducible Feynman diagrams with two external legs and one external double leg
\be
\Gamma^{(2,1)ij}_{\mathrm{B}}\le(\mathbf{k}_1,\mathbf{k}_2\ri)=\Gamma^{\M}_{\mathrm{B}}\le(\mathbf{k}_1,\mathbf{k}_2\ri) \delta^{ij} \, .
\ee
These diagrams can be obtained from those in Fig.~\ref{FeynmanCubic} by replacing one of their three external legs by a double leg.
As before, this amplitude can be formally expanded in a series
\be
\Gamma^{\M}_{\mathrm{B}}=1+\sum_{m=0}^2 h^{\M}_{2-m, m} \le(u_{\rm B}^{\rm I}\ri)^{2-m}\le(u_{\rm B}^{\rm II}\ri)^{m}+\sum_{m=0}^4 h^{\M}_{4-m, m}\le(u_{\rm B}^{\rm I}\ri)^{4-m}\le(u_{\rm B}^{\rm II}\ri)^{m}+O(u_{\rm B}^6)\, 
\ee
with coefficients explicitly calculated as
\be
\sum_{m=0}^2 h^{\M}_{2-m, m} \le(u_{\rm B}^{\rm I}\ri)^{2-m}\le(u_{\rm B}^{\rm II}\ri)^{m}=A^{(3)}_1\le(\sum_{\T_1,\T_2\in\le\{\mathrm{I},\mathrm{II}\ri\}}S_{\T_1,\T_2}u_{\rm B}^{\T_1}u_{\rm B}^{\T_2}\ri)\, 
\ee
and
\bea
\sum_{m=0}^4 h^{\M}_{4-m, m}\le(u_{\rm B}^{\rm I}\ri)^{4-m}\le(u_{\rm B}^{\rm II}\ri)^{m}&=&\le(\frac{3}{2}A^{(3)}_2+B^{(3)}_2\ri) \le(\sum_{\T_1,\T_2\in\le\{\mathrm{I},\mathrm{II}\ri\}}S_{\T_1,\T_2}u_{\rm B}^{\T_1}u_{\rm B}^{\T_2}\ri)^2\\
&&+\le(2B^{(3)}_2+\frac{1}{2}C^{(3)}_2\ri) \le(\sum_{\T_1,\T_2,\T_3,\T_4,\T_5\in\le\{\mathrm{I},\mathrm{II}\ri\}}S_{\T_1,\T_5}a^{\T_5}_{\T_2,\T_3,\T_4}u_{\rm B}^{\T_1}u_{\rm B}^{\T_2}u_{\rm B}^{\T_3}u_{\rm B}^{\T_4}\ri)\, .\nonumber
\eea
We renormalize the bare amplitude by introducing another renormalization factor
\be
Z_{\phi^2}=1+\sum_{m=0}^2 f^{\M}_{2-m, m} \le(g_{\rm P}^{\rm I}\ri)^{2-m}\le(g_{\rm P}^{\rm II}\ri)^{m}+\sum_{m=0}^4 f^{\M}_{4-m, m} \le(g_{\rm P}^{\rm I}\ri)^{4-m}\le(g_{\rm P}^{\rm II}\ri)^{m}+O(g_{\rm P}^6)\, ,
\ee
and requiring that
\be\label{R4}
\Gamma^{\M}_{\mathrm{P}}\le(\mathbf{k}_1,\mathbf{k}_2\ri)\equiv Z_{\phi^2}\le[\Gamma^{\M}_{\mathrm{B}}\le(\mathbf{k}_1,\mathbf{k}_2\ri)\ri]\, 
\ee
remains finite. This condition yields at one-loop
\bea
\sum_{m=0}^2 f^{\M}_{2-m, m} \le(g_{\rm P}^{\rm I}\ri)^{2-m}\le(g_{\rm P}^{\rm II}\ri)^{m}&=&-\sum_{m=0}^2 [\![K_d^{-1}h^{\M}_{2-m, m}]\!]_{\rm s} \le(g_{\rm P}^{\rm I}\ri)^{2-m}\le(g_{\rm P}^{\rm II}\ri)^{m}\\
&=&\le(\frac{-1}{\epsilon}\ri)\le(\sum_{\T_1,\T_2\in\le\{\mathrm{I},\mathrm{II}\ri\}}S_{\T_1,\T_2}g_{\rm P}^{\T_1}g_{\rm P}^{\T_2}\ri)\, ,\nonumber
\eea
and at two-loop%there were typoes in ver1, and $m_2$ should have run from $-1$ to $3$, NOT from $0$ to $3$.
\bea
\sum_{m=0}^4 f^{\M}_{4-m, m} \le(g_{\rm P}^{\rm I}\ri)^{4-m}\le(g_{\rm P}^{\rm II}\ri)^{m}&=&\frac{1}{4\epsilon^2}\le(1+\frac{1}{12}\epsilon\ri)\le(\sum_{\T_1,\T_2\in\le\{\mathrm{I},\mathrm{II}\ri\}}S_{\T_1,\T_2}g_{\rm P}^{\T_1}g_{\rm P}^{\T_2}\ri)^2\\
&&+\frac{1}{\epsilon^2}\le(1-\frac{1}{2}\epsilon\ri) \le(\sum_{\T_1,\T_2,\T_3,\T_4,\T_5\in\le\{\mathrm{I},\mathrm{II}\ri\}}S_{\T_1,\T_5}a^{\T_5}_{\T_2,\T_3,\T_4}g_{\rm P}^{\T_1}g_{\rm P}^{\T_2}g_{\rm P}^{\T_3}g_{\rm P}^{\T_4}\ri)\, .\nonumber
\eea
The critical exponents can then be obtained through the relations
\be
\eta=\gamma^{(\phi)}
\ee
and
\be
\nu^{-1}-2=\gamma^{(\phi^2)}-\eta\, ,
\ee
where the anomalous scaling dimensions are defined by
\be
\gamma^{(\phi)}\equiv\mu\frac{\partial \mathrm{log}\le(Z_{\phi}\ri)}{\partial \mu}=\sum_{m=0}^2 \gamma^{(\phi)}_{2-m, m} \le(g_{\rm P}^{\rm I}\ri)^{2-m}\le(g_{\rm P}^{\rm II}\ri)^{m}+\sum_{m=0}^4 \gamma^{(\phi)}_{4-m, m} \le(g_{\rm P}^{\rm I}\ri)^{4-m}\le(g_{\rm P}^{\rm II}\ri)^{m}+O(g_{\rm P}^6)\, 
\ee
and
\be
\gamma^{(\phi^2)}\equiv\mu\frac{\partial \mathrm{log}\le(Z_{\phi^2}\ri)}{\partial \mu}=\sum_{m=0}^2 \gamma^{(\phi^2)}_{2-m, m} \le(g_{\rm P}^{\rm I}\ri)^{2-m}\le(g_{\rm P}^{\rm II}\ri)^{m}+\sum_{m=0}^4 \gamma^{(\phi^2)}_{4-m, m} \le(g_{\rm P}^{\rm I}\ri)^{4-m}\le(g_{\rm P}^{\rm II}\ri)^{m}+O(g_{\rm P}^6)\, .
\ee
Explicitly we obtain
\be
\sum_{m=0}^2 \left[ {\begin{array}{cc}
\gamma^{(\phi)}_{2-m, m} \, , & \gamma^{(\phi^2)}_{2-m, m}
  \end{array} } \right]\le(g_{\rm P}^{\rm I}\ri)^{2-m}\le(g_{\rm P}^{\rm II}\ri)^{m}=\left[ {\begin{array}{cc}
   \frac{1}{6}\, , & 1
  \end{array} } \right] \le(\sum_{\T_1,\T_2\in\le\{\mathrm{I},\mathrm{II}\ri\}}S_{\T_1,\T_2}g_{\rm P}^{\T_1}g_{\rm P}^{\T_2}\ri)\, 
\ee
and
\bea
\sum_{m=0}^4 \left[ {\begin{array}{cc}
\gamma^{(\phi)}_{4-m, m} \, , & \gamma^{(\phi^2)}_{4-m, m}
  \end{array} } \right]\le(g_{\rm P}^{\rm I}\ri)^{4-m}\le(g_{\rm P}^{\rm II}\ri)^{m}&=&\left[ {\begin{array}{cc}
   -\frac{11}{216}\, , & -\frac{1}{24}
  \end{array} } \right]\le(\sum_{\T_1,\T_2\in\le\{\mathrm{I},\mathrm{II}\ri\}}S_{\T_1,\T_2}g_{\rm P}^{\T_1}g_{\rm P}^{\T_2}\ri)^2\\
  &&+\left[ {\begin{array}{cc}
  \frac{1}{9}\, , & 1
  \end{array} } \right]\le(\sum_{\T_1,\T_2,\T_3,\T_4,\T_5\in\le\{\mathrm{I},\mathrm{II}\ri\}}S_{\T_1,\T_5}a^{\T_5}_{\T_2,\T_3,\T_4}g_{\rm P}^{\T_1}g_{\rm P}^{\T_2}g_{\rm P}^{\T_3}g_{\rm P}^{\T_4}\ri)\, .\nonumber
\eea

\section{Resummed renormalization group equations}\label{Bresum3}
In this section we present three-loop $\beta$-functions and critical exponents for the replica field theory within the dimensional regularization scheme, study the large-order behavior of the perturbative series, and then use these results to resum the series.

\floatsetup[table]{capposition=top}
\begin{table}[b]
\caption{Translation tables for (left) self-energy and (right) cubic-vertex contributions up to three-loop order}
\begin{tabular}{cc}%
\begin{tabular}[t]{|c|c|}
\hline
\textbf{Ref.~\onlinecite{BKM81}} & \textbf{Here} \\
\hline
\hline
$\alpha g_{\rm R}^2$ & $I_2(g_{\rm P})$  \\
\hline
$\alpha\beta g_{\rm R}^4$ & $I_4(g_{\rm P})$  \\
\hline
$\alpha^2 g_{\rm R}^4$ &  $I_2^2(g_{\rm P})$ \\
\hline
$\alpha\gamma g_{\rm R}^6$ &  $I_{6,A}(g_{\rm P})$ \\
\hline
$\alpha\beta^2 g_{\rm R}^6$ &  $I_{6,B}(g_{\rm P})$ \\
\hline
$\alpha^2\beta g_{\rm R}^6$ &  $I_2(g_{\rm P})I_4(g_{\rm P})$ \\
\hline
$\alpha^3 g_{\rm R}^6$ &  $I_2^3(g_{\rm P})$ \\
\hline
\end{tabular}
 &
\begin{tabular}[t]{|c|c|}
\hline
\textbf{Ref.~\onlinecite{BKM81}} & \textbf{Here} \\
\hline
\hline
$\beta g_{\rm R}^3$ & $I_3^{\T}(g_{\rm P})$  \\
\hline
$\gamma g_{\rm R}^5$ & $I_{5,A}^{\T}(g_{\rm P})$  \\
\hline
$\beta^2 g_{\rm R}^5$ &  $I_{5,B}^{\T}(g_{\rm P})$ \\
\hline
$\alpha\beta g_{\rm R}^5$ &  $I_2(g_{\rm P})I_{3}^{\T}(g_{\rm P})$ \\
\hline
$\delta g_{\rm R}^7$ &  $I_{7,A}^{\T}(g_{\rm P})$ \\
\hline
$\lambda g_{\rm R}^7$ &  $I_{7,B}^{\T}(g_{\rm P})$ \\
\hline
$\beta^3 g_{\rm R}^7$ &  $I_{7,C}^{\T}(g_{\rm P})$\ \ \ or\ \ \ $I_{7,D}^{\T}(g_{\rm P})$ \\
\hline
$\beta\gamma g_{\rm R}^7$ &  $I_{7,E}^{\T}(g_{\rm P})$\ \ \ or\ \ \ $I_{7,F}^{\T}(g_{\rm P})$\ \ \ or\ \ \ $I_{7,G}^{\T}(g_{\rm P})$ \\
\hline
$\alpha\gamma g_{\rm R}^7$ &  $I_2(g_{\rm P})I_{5,A}^{\T}(g_{\rm P})$ \\
\hline
$\alpha\beta^2 g_{\rm R}^7$ &  $I_2(g_{\rm P})I_{5,B}^{\T}(g_{\rm P})$\ \ \ or\ \ \ $I_4(g_{\rm P})I_{3}^{\T}(g_{\rm P})$ \\
\hline
$\alpha^2\beta g_{\rm R}^7$ &  $I_2^2(g_{\rm P})I_{3}^{\T}(g_{\rm P})$ \\
\hline
\end{tabular} \tabularnewline
\end{tabular}
\label{SelfTranslation}
\end{table}

\subsection{Three-loop perturbative expressions}
In order to concisely display the results, let us first define:
\bea
I_2(g_{\rm P})&\equiv&\sum_{\T_1,\T_2\in\le\{\mathrm{I},\mathrm{II}\ri\}}S_{\T_1,\T_2}g_{\rm P}^{\T_1}g_{\rm P}^{\T_2}\, ,\label{simple1}\\
I^{\T}_{3}(g_{\rm P})&\equiv&\sum_{\T_1,\T_2,\T_3\in\le\{\mathrm{I},\mathrm{II}\ri\}}a^{\T}_{\T_1,\T_2,\T_3}g_{\rm P}^{\T_1}g_{\rm P}^{\T_2}g_{\rm P}^{\T_3}\, ,\\
I_{4}(g_{\rm P})&\equiv&\sum_{\T_1,\T_2,\T_3,\T_4,\T_5\in\le\{\mathrm{I},\mathrm{II}\ri\}}S_{\T_1,\T_5}a^{\T_5}_{\T_2,\T_3,\T_4}g_{\rm P}^{\T_1}g_{\rm P}^{\T_2}g_{\rm P}^{\T_3}g_{\rm P}^{\T_4}\, ,\\
I^{\T}_{5,A}(g_{\rm P})&\equiv&\sum_{\T_1,\T_2,\T_3,\T_4,\T_5\in\le\{\mathrm{I},\mathrm{II}\ri\}}a^{\T}_{\T_1,\T_2,\T_3;\T_4,\T_5}g_{\rm P}^{\T_1}g_{\rm P}^{\T_2}g_{\rm P}^{\T_3}g_{\rm P}^{\T_4}g_{\rm P}^{\T_5}\, ,\\
I^{\T}_{5,B}(g_{\rm P})&\equiv&\sum_{\T_1,\T_2,\T_3,\T_4,\T_5,\T_6\in\le\{\mathrm{I},\mathrm{II}\ri\}}a^{\T}_{\T_1,\T_2,\T_6}a^{\T_6}_{\T_3,\T_4,\T_5}g_{\rm P}^{\T_1}g_{\rm P}^{\T_2}g_{\rm P}^{\T_3}g_{\rm P}^{\T_4}g_{\rm P}^{\T_5}\, ,\label{simple5}\\
I_{6,A}(g_{\rm P})&\equiv&\sum_{\T_1,\T_2,\T_3,\T_4,\T_5,\T_6,\T_7\in\le\{\mathrm{I},\mathrm{II}\ri\}}S_{\T_1,\T_7}a^{\T_7}_{\T_2,\T_3,\T_4; \T_5,\T_6}g_{\rm P}^{\T_1}g_{\rm P}^{\T_2}g_{\rm P}^{\T_3}g_{\rm P}^{\T_4}g_{\rm P}^{\T_5}g_{\rm P}^{\T_6}\, ,\\
I_{6,B}(g_{\rm P})&\equiv&\sum_{\T_1,\T_2,\T_3,\T_4,\T_5,\T_6,\T_7,\T_8\in\le\{\mathrm{I},\mathrm{II}\ri\}}S_{\T_1,\T_7}a^{\T_7}_{\T_2,\T_3,\T_8}a^{\T_8}_{\T_4,\T_5,\T_6}g_{\rm P}^{\T_1}g_{\rm P}^{\T_2}g_{\rm P}^{\T_3}g_{\rm P}^{\T_4}g_{\rm P}^{\T_5}g_{\rm P}^{\T_6}\, ,\\
I^{\T}_{7,A}(g_{\rm P})&\equiv&\sum_{\T_1,\T_2,\T_3,\T_4,\T_5,\T_6,\T_7\in\le\{\mathrm{I},\mathrm{II}\ri\}}a^{\T}_{\T_1,\T_2,\T_3,\T_4,\T_5,\T_6,\T_7}g_{\rm P}^{\T_1}g_{\rm P}^{\T_2}g_{\rm P}^{\T_3}g_{\rm P}^{\T_4}g_{\rm P}^{\T_5}g_{\rm P}^{\T_6}g_{\rm P}^{\T_7}\, ,\\
I^{\T}_{7,B}(g_{\rm P})&\equiv&\sum_{\T_1,\T_2,\T_3,\T_4,\T_5,\T_6,\T_7\in\le\{\mathrm{I},\mathrm{II}\ri\}}b^{\T}_{\T_1,\T_2,\T_3,\T_4,\T_5,\T_6,\T_7}g_{\rm P}^{\T_1}g_{\rm P}^{\T_2}g_{\rm P}^{\T_3}g_{\rm P}^{\T_4}g_{\rm P}^{\T_5}g_{\rm P}^{\T_6}g_{\rm P}^{\T_7}\, ,\\
I^{\T}_{7,C}(g_{\rm P})&\equiv&\sum_{\T_1,\T_2,\T_3,\T_4,\T_5,\T_6,\T_7,\T_8,\T_9\in\le\{\mathrm{I},\mathrm{II}\ri\}}a^{\T}_{\T_1,\T_2,\T_8}a^{\T_8}_{\T_3,\T_4,\T_9}a^{\T_9}_{\T_5,\T_6,\T_7}g_{\rm P}^{\T_1}g_{\rm P}^{\T_2}g_{\rm P}^{\T_3}g_{\rm P}^{\T_4}g_{\rm P}^{\T_5}g_{\rm P}^{\T_6}g_{\rm P}^{\T_7}\, ,\\
I^{\T}_{7,D}(g_{\rm P})&\equiv&\sum_{\T_1,\T_2,\T_3,\T_4,\T_5,\T_6,\T_7,\T_8,\T_9\in\le\{\mathrm{I},\mathrm{II}\ri\}}a^{\T}_{\T_1,\T_8,\T_9}a^{\T_8}_{\T_2,\T_3,\T_4}a^{\T_9}_{\T_5,\T_6,\T_7}g_{\rm P}^{\T_1}g_{\rm P}^{\T_2}g_{\rm P}^{\T_3}g_{\rm P}^{\T_4}g_{\rm P}^{\T_5}g_{\rm P}^{\T_6}g_{\rm P}^{\T_7}\, ,\\
I^{\T}_{7,E}(g_{\rm P})&\equiv&\sum_{\T_1,\T_2,\T_3,\T_4,\T_5,\T_6,\T_7,\T_8\in\le\{\mathrm{I},\mathrm{II}\ri\}}a^{\T}_{\T_1,\T_2,\T_8;\T_3,\T_4}a^{\T_8}_{\T_5,\T_6,\T_7}g_{\rm P}^{\T_1}g_{\rm P}^{\T_2}g_{\rm P}^{\T_3}g_{\rm P}^{\T_4}g_{\rm P}^{\T_5}g_{\rm P}^{\T_6}g_{\rm P}^{\T_7}\, ,\\
I^{\T}_{7,F}(g_{\rm P})&\equiv&\sum_{\T_1,\T_2,\T_3,\T_4,\T_5,\T_6,\T_7,\T_8\in\le\{\mathrm{I},\mathrm{II}\ri\}}a^{\T}_{\T_1,\T_2,\T_3;\T_4,\T_8}a^{\T_8}_{\T_5,\T_6,\T_7}g_{\rm P}^{\T_1}g_{\rm P}^{\T_2}g_{\rm P}^{\T_3}g_{\rm P}^{\T_4}g_{\rm P}^{\T_5}g_{\rm P}^{\T_6}g_{\rm P}^{\T_7}\, ,\\
I^{\T}_{7,G}(g_{\rm P})&\equiv&\sum_{\T_1,\T_2,\T_3,\T_4,\T_5,\T_6,\T_7,\T_8\in\le\{\mathrm{I},\mathrm{II}\ri\}}a^{\T}_{\T_1,\T_2,\T_8}a^{\T_8}_{\T_3,\T_4,\T_5;\T_6,\T_7}g_{\rm P}^{\T_1}g_{\rm P}^{\T_2}g_{\rm P}^{\T_3}g_{\rm P}^{\T_4}g_{\rm P}^{\T_5}g_{\rm P}^{\T_6}g_{\rm P}^{\T_7}\, ,
\eea
where we introduced two new combinatorial factors through
\be
\label{definea7}
\sum_{i_4,i_5,i_6,i_7,i_8,i_9,i_{10},i_{11},i_{12}=1}^{\frac{n(n-3)}{2}}T_{\T_1}^{i_1 i_4 i_5}T_{\T_2}^{i_2 i_6 i_7}T_{\T_3}^{i_3 i_8 i_9}T_{\T_4}^{i_4 i_6 i_{10}}T_{\T_5}^{i_5 i_8 i_{11}}T_{\T_6}^{i_7 i_9 i_{12}}T_{\T_7}^{i_{10} i_{11} i_{12}}
\equiv \sum_{\T\in\le\{\mathrm{I},\mathrm{II}\ri\}}a^{\T}_{\T_1,\T_2,\T_3,\T_4,\T_5,\T_6,\T_7} T_{\T}^{i_1 i_2 i_3}\, 
\ee
and
\be
\label{defineb7}
\sum_{i_4,i_5,i_6,i_7,i_8,i_9,i_{10},i_{11},i_{12}=1}^{\frac{n(n-3)}{2}}T_{\T_1}^{i_1 i_4 i_5}T_{\T_2}^{i_2 i_6 i_7}T_{\T_3}^{i_3 i_8 i_9}T_{\T_4}^{i_4 i_6 i_{10}}T_{\T_5}^{i_5 i_8 i_{11}}T_{\T_6}^{i_7 i_{11} i_{12}}T_{\T_7}^{i_9 i_{10} i_{12}}
\equiv \sum_{\T\in\le\{\mathrm{I},\mathrm{II}\ri\}}b^{\T}_{\T_1,\T_2,\T_3,\T_4,\T_5,\T_6,\T_7} T_{\T}^{i_1 i_2 i_3}\, 
\ee

As alluded to before, three-loop results were obtained in Ref.~\cite{BKM81} for simpler theories with one cubic coupling.
We need here to generalize their results to a theory with multiple cubic couplings.
Looking at Feynman diagrams and corresponding contributions in Tables~A1 and A2 of Ref.~\cite{BKM81}, we arrive at the dictionary in Table~\ref{SelfTranslation}. Note that, for three-loop $\beta$-functions, some terms subdivide into a few distinct possibilities; relative ratio can be obtained by reading off the $\epsilon^{-1}$-term in the corresponding amplitudes. With such a dictionary, we obtain critical exponents [recall $\eta=\gamma^{(\phi)}$ and $\nu^{-1}-2=\gamma^{(\phi^2)}-\eta$]
\bea\label{3loopgamma}
\gamma^{(\phi)}&=&\frac{1}{6}I_2-\frac{11}{216}I_2^2+\frac{1}{9}I_4+\frac{821}{31104}I_2^3-\frac{179}{1728}I_2I_4+\le\{\frac{7}{48}-\frac{\zeta(3)}{12}\ri\}I_{6,A}+\frac{85}{864}I_{6,B}\, ,\\
\gamma^{(\phi^2)}&=&I_2-\frac{1}{24}I_2^2+I_4+\frac{95}{216}I_2^3-\le\{\frac{79}{96}+\frac{\zeta(3)}{2}\ri\}I_2I_4+\frac{7}{8}I_{6,A}+\le\{\frac{65}{48}+\zeta(3)\ri\}I_{6,B}\, ,
\eea
and $\beta$-functions
\bea
\beta^{\T}&=&\le(-\frac{\epsilon}{2}+\frac{3}{2}\eta\ri)g_{\rm P}^{\T}-I_{3}^{\T}+\frac{7}{24}I_2I_3^{\T}-\frac{1}{2}I_{5,A}^{\T}-\frac{3}{4}I_{5,B}^{\T}\label{3loopbeta}\\
&&-\frac{119}{864}I_2^2I_3^{\T}+\frac{11}{48}I_2I_{5,A}^{\T}+\frac{7}{32}I_2I_{5,B}^{\T}+\frac{23}{96}I_4I_3^{\T}\nonumber\\
&&-I_{7,A}^{\T}+\le\{1-3\zeta(3)\ri\}I_{7,B}^{\T}-\frac{3}{8}I_{7,C}^{\T}+\frac{15}{16}I_{7,D}^{\T}-\frac{3}{16}I_{7,E}^{\T}+\le\{-\frac{23}{24}+\zeta(3)\ri\}I_{7,F}^{\T}+\le\{-\frac{29}{16}+\frac{3}{2}\zeta(3)\ri\}I_{7,G}^{\T}\, ,\nonumber
\eea
where $\zeta(3)\equiv\sum_{n=1}^{\infty}\frac{1}{n^3}$. Note that when truncated to two-loop order, they reproduce the two-loop results, as they should.

For the replica field theory, the most demanding part of higher-loop calculations is evaluating the combinatorial factors, defined in Eqs.~\eqref{definea7}) and \eqref{defineb7} for those arising at three-loop order. We evaluated them using the method described in Sec.~\ref{realhorror}, here fitting combinatorial factors obtained for $n=6,\ldots,21$ by functions $\frac{c_0+c_1n+\ldots+c_{15}n^{15}}{512(n-1)^3 (n-2)^9}$ with integer coefficients $c_0,\ldots,c_{15}$, and then cross-validating the consistency of the results against values obtained for $n=22$. These evaluations require quadruple numerical precision. 
See Supplemental Material at http://dx.doi.org/10.7924/G86Q1V5C for the results. We further checked that the three-loop combinatorial factors thus obtained satisfy the nontrivial 't Hooft identities
\be
\sum_{\T_9\in\le\{\mathrm{I},\mathrm{II}\ri\}}S_{\T_1,\T_9}a^{\T_9}_{\T_2,\T_3,\T_4,\T_5,\T_6,\T_7,\T_8}=\sum_{\T_9\in\le\{\mathrm{I},\mathrm{II}\ri\}}S_{\T_2,\T_9}a^{\T_9}_{\T_1,\T_5,\T_6,\T_3,\T_4,\T_8,\T_7}\, 
\ee
and
\be
\sum_{\T_9\in\le\{\mathrm{I},\mathrm{II}\ri\}}S_{\T_1,\T_9}b^{\T_9}_{\T_2,\T_3,\T_4,\T_5,\T_6,\T_7,\T_8}=\sum_{\T_9\in\le\{\mathrm{I},\mathrm{II}\ri\}}S_{\T_2,\T_9}b^{\T_9}_{\T_1,\T_5,\T_6,\T_3,\T_4,\T_8,\T_7}\, .
\ee

\subsection{Large-order behavior}
In order to derive the equations of motion for the replicon field, we introduce Lagrange multiplier fields $\lambda_{a}\le(\mathbf{x}\ri)$ for the replicon constraint equations $\sum_{b=1}^n\phi_{ab}\le(\mathbf{x}\ri)=0$~\cite{Yaida16}, and extremize
\begin{equation*}
I[\phi_{ab}\le(\mathbf{x}\ri),\lambda_{a}\le(\mathbf{x}\ri)]=\int\!\! \mathrm{d}\mathbf{x}\le\{\frac{1}{2}\sum_{a,b=1}^{n}\le(\nabla\phi_{ab}\ri)^2+\frac{\mu^2}{2}\sum_{a,b=1}^{n}\phi_{ab}^2-\frac{1}{3!}\le(g_{\rm bare}^{\rm I}\!\!\sum_{a,b=1}^n\!\!\phi_{ab}^3+g_{\rm bare}^{\rm II}\!\!\sum_{a,b,c=1}^n\!\!\phi_{ab}\phi_{bc}\phi_{ca}\ri)-\sum_{a,b=1}^n\lambda_{a}\phi_{ab}\ri\}\, .
\end{equation*}
We also include in the expression a quadratic mass term, $\mu^2$, which gives a scale to the problem and enables the large-order analysis away from the upper critical dimension~\cite{BP78}.
Field variations give the replicon equations of motion for $a\ne b$,
\be
(-\nabla^2+\mu^2) \phi_{ab}-\frac{1}{2}\le[g_{\rm bare}^{\rm I} \phi_{ab}^2+g_{\rm bare}^{\rm II} \sum_{c=1}^n\phi_{ac}\phi_{cb}\ri]=\frac{\lambda_{a}+\lambda_{b}}{2}\, .
\ee
At large order, the bare coupling can be substituted by the physical couplings through the tree-level relation~\cite{BP78}, \textit{i.e.}, $K_{d}^{\frac{1}{2}}\mu^{-\frac{\epsilon}{2}}(g_{\mathrm{bare}}^{\mathrm{I}},g_{\mathrm{bare}}^{\mathrm{II}})\approx(g_{\rm P}^{\rm I},g_{\rm P}^{\rm II})\equiv g_{\rm P}(\cos\theta,\sin\theta)$ [\textit{cf}.~Eqs.~\eqref{dimlessbare}, \eqref{sphere}, and \eqref{PBrelation}], which leads to
\be
(-\nabla^2+\mu^2) \phi_{ab}-\frac{g_{\rm P}K_{d}^{-\frac{1}{2}}\mu^{\frac{\epsilon}{2}}}{2}\le[\cos\theta \phi_{ab}^2+\sin\theta \sum_{c=1}^n\phi_{ac}\phi_{cb}\ri]=\frac{\lambda_{a}+\lambda_{b}}{2}\, .
\ee

As is standard~\cite{YMA05} and proven in some case~\cite{CGM78}, we assume that the saddle-point solution governing the large-order behavior takes the separable and spherically-symmetric form
\bea
\phi^{\star}_{ab}\le(\mathbf{x}\ri)&=&\frac{K_{d}^{\frac{1}{2}}\mu^{-\frac{\epsilon}{2}}}{g_{\rm P}}\mu^2 \tilde{F}\le(|\mu\mathbf{x}|\ri) v_{ab}\, \\
\lambda^{\star}_{a}\le(\mathbf{x}\ri)&=&\frac{K_{d}^{\frac{1}{2}}\mu^{-\frac{\epsilon}{2}}}{g_{\rm P}}\mu^4 \tilde{F}^2\le(|\mu\mathbf{x}|\ri) w_{a}\, .
\eea
Here the dimensionless spherical function $\tilde{F}\le(\tilde{r}\ri)$ satisfies
\be\label{defF}
\le[\frac{\mathrm{d}^2}{\mathrm{d}\tilde{r}^2}+\frac{(d-1)}{\tilde{r}}\frac{\mathrm{d}}{\mathrm{d}\tilde{r}}\ri]\tilde{F}=\tilde{F}-\tilde{F}^2\, \ \ \ \mathrm{and}\ \ \ \frac{\mathrm{d}F}{\mathrm{d}\tilde{r}}\Bigg|_{\tilde{r}=0}=0\, ,
\ee
and the constant symmetric matrix $v_{ab}$ satisfies the replicon constraints
\be\label{RC}
\sum_{b=1}^nv_{ab}=0\, \ \ \ \mathrm{for}\ \ \ a=1,\ldots,n\, .
\ee
This ansatz solves the replicon equations of motion for $a\ne b$ if and only if
\be\label{ME}
v_{ab}-\frac{1}{2}\le[\cos\theta v_{ab}^2+\sin\theta \sum_{c=1}^nv_{ac}v_{cb}\ri]=\frac{w_{a}+w_{b}}{2}\, .
\ee

For general spatial dimensions $d$, the spherically symmetric function $\tilde{F}\le(\tilde{r}\ri)$ can be obtained numerically by the pseudospectral method, as described in subsection~\ref{sec:pseudospectral}, whereas the matrix equation \eqref{ME} can be solved analytically by adapting the 1-step RSB ansatz~\cite{Parisi79,MPV87,CC05,Denef11}.
Specifically, by relabeling the replica index as  $a=(\Pa_{0}-1)m_1+\Pa_{1}$, with $\Pa_0=1,\ldots,\frac{n}{m_1}$ specifying the cluster of metastable states and $\Pa_1=1,\ldots,m_1$ the state within that cluster, we obtain
\be
v_{ab}=v_0(1-\delta_{\Pa_0,\Pb_0})+v_1\delta_{\Pa_0,\Pb_0}(1-\delta_{\Pa_1,\Pb_1})\, .
\ee
The only constant vector $w_{a}$ compatible with this ansatz is $w_{a}=w$, independently of the replica index. The matrix equation \eqref{ME} under the replicon constraints \eqref{RC} yields in the replica limit $n\rightarrow0$ (where $0\leq m_1\leq1$)
\bea
v_0&=&(1-m_1)\le(\frac{2}{\cos\theta}\ri)\le(\frac{1}{1-2m_1+m_1\tan\theta}\ri)\\
v_1&=&-m_1\le(\frac{2}{\cos\theta}\ri)\le(\frac{1}{1-2m_1+m_1\tan\theta}\ri)\\
w&=&-m_1(1-m_1)\le(\frac{2}{\cos\theta}\ri)\le(\frac{1}{1-2m_1+m_1\tan\theta}\ri)^2(1-\tan\theta)\, .
\eea
The saddle-point replicon action
\bea
\lim_{n\rightarrow0}\frac{S[\phi^{\star}_{ab}\le(\mathbf{x}\ri)]}{n}&=&\lim_{n\rightarrow0}\frac{I[\phi^{\star}_{ab}\le(\mathbf{x}\ri),\lambda^{\star}_{a}\le(\mathbf{x}\ri)]}{n}\\
&=&\frac{(2\pi)^d K_d^2}{g_P^2}\le[\int_0^{\infty} \mathrm{d}\tilde{r} \tilde{r}^{d-1}\tilde{F}^3\ri]\lim_{n\rightarrow0}\le[\frac{\frac{1}{2}\sum_{a,b=1}^{n}v_{ab}^2-\frac{1}{3!}\le(\cos\theta\sum_{a,b=1}^nv_{ab}^3+\sin\theta\sum_{a,b,c=1}^nv_{ab}v_{bc}v_{ca}\ri)}{n}\ri]\nonumber\\
&=&-\frac{1}{6}\frac{(2\pi)^d K_d^2}{g_P^2}\le[\int_0^{\infty} \mathrm{d}\tilde{r} \tilde{r}^{d-1} \tilde{F}^3\ri]\le(\frac{2}{\cos\theta}\ri)^2\le[\frac{m_1(1-m_1)}{\le(1-2m_1+m_1 \tan\theta\ri)^2}\ri]\, ,\nonumber
\eea
where
\be
\int_0^{\infty} \mathrm{d}\tilde{r} \tilde{r}^{d-1}\le[\le(\frac{\mathrm{d}\tilde{F}}{\mathrm{d}\tilde{r}}\ri)^2+\tilde{F}^2\ri]=\int_0^{\infty} \mathrm{d}\tilde{r} \tilde{r}^{d-1} \tilde{F}^3\, 
\ee
follows from integrating Eq.~\eqref{defF} by parts. Note that the regularity at the origin $\frac{\mathrm{d}\tilde{F}}{\mathrm{d}\tilde{r}}\Big|_{\tilde{r}=0}=0$, and the proper decay at $\tilde{r}=\infty$ cancel the boundary term. Extremizing the action with respect to the RSB parameter $m_1\in [0,1]$ gives
\be
m_1^{\star}=\frac{1}{\tan\theta}\, .
\ee
The RSB solution thus exists if and only if $1<\tan\theta=\frac{g_{\rm P}^{\rm II}}{g_{\rm P}^{\rm I}}<\infty$, which is completely consistent with the existential condition of the RSB transition at the mean-field level~\cite{FPR96} and nearly coincides with the two-loop basin of attraction. The value of $A(\theta)$ that governs the asymptotic series coefficients $f_{k}(\theta)\sim k! \le[-1/A(\theta)\ri]^k$ at large loop order $k$ is further given by
\be
A(\theta)=-g_P^2\lim_{n\rightarrow0}\frac{S[\phi^{\star}_{ab}\le(\mathbf{x}\ri)]}{n}=\le[\frac{(2\pi)^d K_d^2}{6}\int_0^{\infty} \mathrm{d}\tilde{r} \tilde{r}^{d-1} \tilde{F}^3\ri]\times\le[\frac{1}{\le(\sin\theta-\cos\theta\ri)\cos\theta}\ri]\, ,
\ee
which is positive in the wedge with $1<\frac{g_{\rm P}^{\rm II}}{g_{\rm P}^{\rm I}}<\infty$, and thus validates the Borel-summability of the perturbative field theory precisely within the wedge that contains the fixed point.

\subsection{Borel resummation with the conformal mapping}
We start from an anomalous dimension $\gamma$ with a double-series expansions of the form
\bea
\gamma&=&\sum_{k_1,k_2=0; k_1+k_2=\mathrm{even}}^{\infty} \gamma_{k_1,k_2}\le(g_{\rm P}^{\rm I}\ri)^{k_1}\le(g_{\rm P}^{\rm II}\ri)^{k_2}\\
&=&\sum_{k=0}^{\infty} g_{\rm P}^{2k}\le[\sum_{k_1=0}^{2k}\gamma_{k_1,2k-k_1}\le(\cos\theta\ri)^{k_1}\le(\sin\theta\ri)^{2k-k_1}\ri]\, \nonumber\\
&\equiv&\sum_{k=0}^{\infty}\Upsilon_{k}\le(\theta\ri)g_{\rm P}^{2k}\, .\nonumber
\eea
In this form, the tree-level contribution vanishes, \textit{i.e.} $\Upsilon_{0}=0$, and the three-loop results yield the first three nontrivial coefficients, $\Upsilon_{1,2,3}\le(\theta\ri)$. Its Borel transform is then
\be
\tilde{\gamma}_{\rm B}\le(g_{\rm P}^2; \theta\ri)\equiv \sum_{k=0}^{\infty}\frac{\Upsilon_{k}\le(\theta\ri)}{k!}g_{\rm P}^{2k},
\ee 
and hence $\gamma=\int_{0}^{\infty}\mathrm{d}t e^{-t}\tilde{\gamma}_{\rm B}(g^2t;\theta)$. In the conformally-related coordinate
\be
u(g_{\rm P}^2;\theta)\equiv \frac{\sqrt{1+\frac{g_{\rm P}^2}{A\le(\theta\ri)}}-1}{\sqrt{1+\frac{g_{\rm P}^2}{A\le(\theta\ri)}}+1}\, 
\ee
the Borel transform is expected to have a radius of convergence of unity~\cite{LZ80}.

Matching the expansion coefficients order by order leads to
\be
\tilde{\gamma}_{\rm B}=4A\Upsilon_{1}u+\le(8A\Upsilon_1+8A^2\Upsilon_2\ri)u^2+\le(12A\Upsilon_1+32A^2\Upsilon_2+\frac{32}{3}A^3\Upsilon_3\ri)u^3+O(u^4)\, .
\ee
Truncating the series at the third order in $u$ and performing the inverse Borel transform, we obtain the three-loop resummed expression
\bea
\gamma&=&\le[4A(\theta)\Upsilon_{1}\le(\theta\ri)\ri]\int_0^{\infty} \mathrm{d}t e^{-t}\le[\frac{\sqrt{1+\frac{g_{\rm P}^2t}{A\le(\theta\ri)}}-1}{\sqrt{1+\frac{g_{\rm P}^2t}{A\le(\theta\ri)}}+1}\ri]\\
&&+\le[8A\le(\theta\ri)\Upsilon_1\le(\theta\ri)+8A^2\le(\theta\ri)\Upsilon_2\le(\theta\ri)\ri]\int_0^{\infty}\mathrm{d}t e^{-t}\le[\frac{\sqrt{1+\frac{g_{\rm P}^2t}{A\le(\theta\ri)}}-1}{\sqrt{1+\frac{g_{\rm P}^2t}{A\le(\theta\ri)}}+1}\ri]^2\nonumber\\
&&+\le[12A\le(\theta\ri)\Upsilon_1\le(\theta\ri)+32A^2\le(\theta\ri)\Upsilon_2\le(\theta\ri)+\frac{32}{3}A^3\le(\theta\ri)\Upsilon_3\le(\theta\ri)\ri]\int_0^{\infty}\mathrm{d}t e^{-t}\le[\frac{\sqrt{1+\frac{g_{\rm P}^2t}{A\le(\theta\ri)}}-1}{\sqrt{1+\frac{g_{\rm P}^2t}{A\le(\theta\ri)}}+1}\ri]^3\, .\nonumber
\eea
Similarly,
\bea
\beta^{\T}&=&\sum_{k_1,k_2=0; k_1+k_2=\mathrm{odd}}^{\infty} \beta^{\T}_{k_1,k_2}\le(g_{\rm P}^{\rm I}\ri)^{k_1}\le(g_{\rm P}^{\rm II}\ri)^{k_2}\\
&=&\sum_{k=0}^{\infty} g_{\rm P}^{2k+1}\le[\sum_{k_1=0}^{2k+1}\beta^{\T}_{k_1,2k+1-k_1}\le(\cos\theta\ri)^{k_1}\le(\sin\theta\ri)^{2k+1-k_1}\ri]\, \nonumber\\
&\equiv&g_{\rm P}\sum_{k=0}^{\infty}\mathcal{B}^{\T}_{k}\le(\theta\ri)g_{\rm P}^{2k}\, ,\nonumber
\eea
or in polar coupling coordinates
\bea
\beta^{(g^2)}&\equiv& \mu\frac{\partial (g_{\rm P}^2)}{\partial \mu}=g^2_{\rm P}\sum_{k=0}^{\infty}\le[2\cos\theta\mathcal{B}^{\rm I}_{k}\le(\theta\ri)+2\sin\theta\mathcal{B}^{\rm II}_{k}\le(\theta\ri)\ri]g_{\rm P}^{2k}\equiv g_{\rm P}^2\le[\sum_{k=0}^{\infty}\mathcal{B}^{(g^2)}_{k}\le(\theta\ri)g_{\rm P}^{2k}\ri]\,\label{radibeta} \\
\beta^{(\tan\theta)}&\equiv& \mu\frac{\partial (\tan\theta)}{\partial \mu}=\sum_{k=0}^{\infty}\le[-\frac{\sin\theta}{\cos^2\theta}\mathcal{B}^{\rm I}_{k}\le(\theta\ri)+\frac{1}{\cos\theta}\mathcal{B}^{\rm II}_{k}\le(\theta\ri)\ri]g_{\rm P}^{2k}\equiv \sum_{k=0}^{\infty}\mathcal{B}^{(\tan\theta)}_{k}\le(\theta\ri)g_{\rm P}^{2k}\, .
\eea
For the angular component, the tree-level contribution vanishes, \textit{i.e.} $\mathcal{B}^{(\tan\theta)}_{0}=0$, whereas the radial component has $\mathcal{B}^{(g^2)}_{0}=d-6$. For the angular $\beta$-function, the three-loop resummed expression is similar to that for $\gamma$ with $\Upsilon_{k}$ replaced by $\mathcal{B}^{(\tan\theta)}_{k}$, while for the radial $\beta$-function
\bea
\beta^{(g^2)}
&=&(d-6)g^2_{\rm P}+g^2_{\rm P}\le[4A(\theta)\mathcal{B}^{(g^2)}_{1}\le(\theta\ri)\ri]\int_0^{\infty} \mathrm{d}t e^{-t}\le[\frac{\sqrt{1+\frac{g_{\rm P}^2t}{A\le(\theta\ri)}}-1}{\sqrt{1+\frac{g_{\rm P}^2t}{A\le(\theta\ri)}}+1}\ri]\\
&&+g^2_{\rm P}\le[8A\le(\theta\ri)\mathcal{B}^{(g^2)}_{1}\le(\theta\ri)+8A^2\le(\theta\ri)\mathcal{B}^{(g^2)}_{2}\le(\theta\ri)\ri]\int_0^{\infty}\mathrm{d}t e^{-t}\le[\frac{\sqrt{1+\frac{g_{\rm P}^2t}{A\le(\theta\ri)}}-1}{\sqrt{1+\frac{g_{\rm P}^2t}{A\le(\theta\ri)}}+1}\ri]^2\nonumber\\
&&+g^2_{\rm P}\le[12A\le(\theta\ri)\mathcal{B}^{(g^2)}_{1}\le(\theta\ri)+32A^2\le(\theta\ri)\mathcal{B}^{(g^2)}_{2}\le(\theta\ri)+\frac{32}{3}A^3\le(\theta\ri)\mathcal{B}^{(g^2)}_{3}\le(\theta\ri)\ri]\int_0^{\infty}\mathrm{d}t e^{-t}\le[\frac{\sqrt{1+\frac{g_{\rm P}^2t}{A\le(\theta\ri)}}-1}{\sqrt{1+\frac{g_{\rm P}^2t}{A\le(\theta\ri)}}+1}\ri]^3\, ,\nonumber\\
\eea
after resumming the series in Eq.~\eqref{radibeta}.

\subsection{Pseudospectral method}
\label{sec:pseudospectral}
In order to efficiently solve the boundary value problem of identifying a nontrivial solution to Eq.~\eqref{defF} with Neumann boundary condition $\frac{\mathrm{d}F}{\mathrm{d}\tilde{r}}\Big|_{\tilde{r}=0}=0$ and Dirichlet boundary condition $\tilde{F}(\tilde{r}=\infty)=0$, we use the pseudospectral method~\cite{SpectralMatlab}.
The function $\tilde{F}(\tilde{r})$ is thus represented in a basis of $N$ Chebyshev polynomials, $T_k(x(\tilde{r}))$, keeping track of the function values at the Chebyshev extrema collocation grid.  We define the coordinate $x(\tilde{r})\equiv b_0\tanh[0.1(\tilde{r}-1)]+b_1$ with $b_0$ and $b_1$ chosen such that the domain $\tilde{r}\in [0,\infty]$ maps onto the compact interval $x\in [-1,1]$. Once the coordinate parameter is established, the nonlinear equations for function values at the collocation points are solved by Newton's method.

The integration result is sensitive to the proximity of the initial guess to the nontrivial saddle-point solution: bad guesses diverge away from it. In order to circumvent this problem, we adopt the mountain pass algorithm developed in Ref.~\cite{AALY15}.
First, the full domain is subdivided in two: $[0,1]$ and $[1,\infty]$. In the first region, the solution to the saddle-point equation with the boundary conditions $\frac{\mathrm{d}F}{\mathrm{d}\tilde{r}}\Big|_{\tilde{r}=0}=0$ and $\tilde{F}(\tilde{r}=1)=\tilde{F}_{\rm M}$ is obtained, while in the second region the solution with $\tilde{F}(\tilde{r}=1)=\tilde{F}_{\rm M}$ and $\tilde{F}(\tilde{r}=\infty)=0$ is obtained. For a generic choice of middle-point value $\tilde{F}_{\rm M}$, the patched solution has a kink at $\tilde{r}=1$. A good initial guess for the smooth solution is attained by varying $\tilde{F}_{\rm M}$ until the left- and right-sided first derivatives match. In order to facilitate this search, the process is bootstrapped. That is, after finding a solution in spatial dimension $d=1$ higher-dimensional solutions are obtained by adiabatically increasing $d$ in steps $\Delta d=0.001$. The resulting $d$-dependent constant (Fig.~\ref{cdinfo})
\be
c_d\equiv\frac{(2\pi)^d K_d^2}{6}\int_0^{\infty} \mathrm{d}\tilde{r} \tilde{r}^{d-1} \tilde{F}^3\, 
\ee
controls the large order behavior, $A(\theta)=c_d/[\cos\theta (\sin\theta-\cos\theta)]$. Note that our results are robust against changes to the number of collocation points, as long as it is sufficiently large.

\begin{figure*}[t]
\centerline{\includegraphics[width=0.33\textwidth]{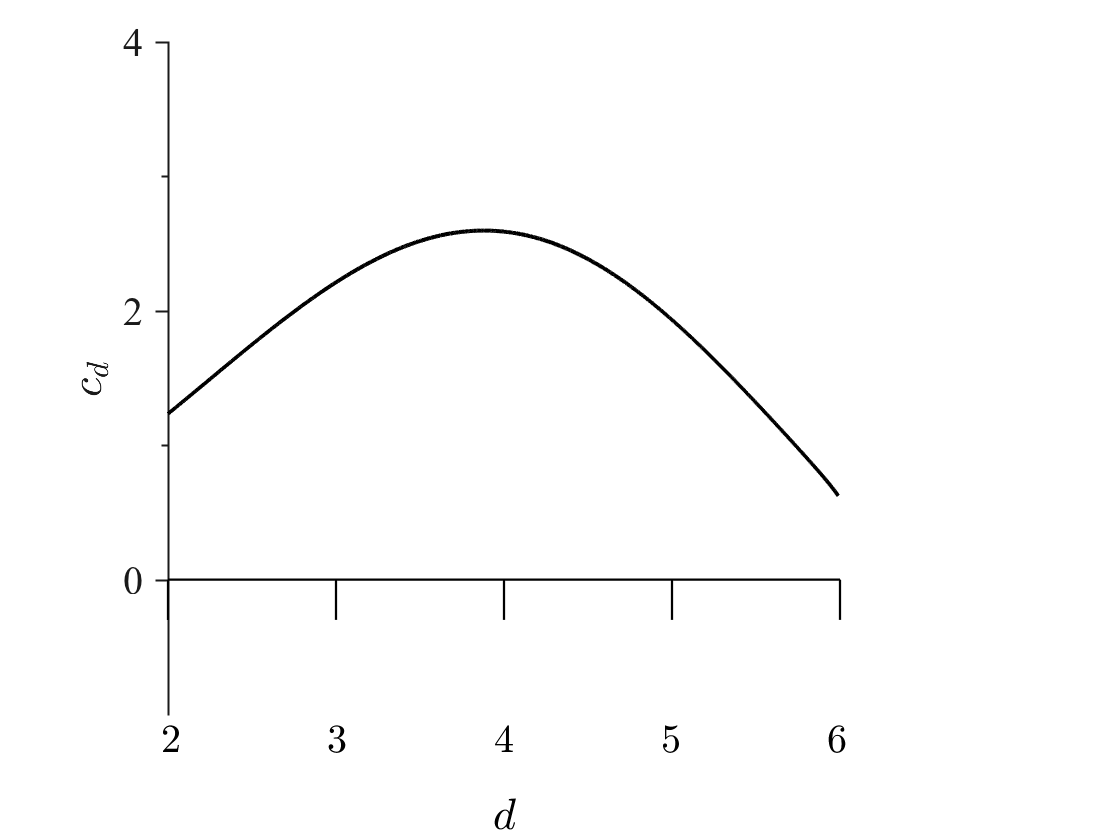}}
\caption{The $d$-dependent constant $c_d$ that governs the large-order behavior of the perturbative series through $A(\theta)=c_d/[\cos\theta (\sin\theta-\cos\theta)]$ is obtained using $100$ collocation points.}
\label{cdinfo}
\end{figure*}

\end{widetext}

\bibliography{RG}

\begin{thebibliography}{75}
\bibitem [{foo({\natexlab{a}})}]{footnote_statefollowing}%
  \BibitemOpen
  \href@noop {} {} {Strictly speaking the
  Gardner transition exists in finite dimensions only in the high-overlap state
  of the $\varepsilon$-quenched Franz-Parisi potential for
  $\varepsilon>\varepsilon_{\rm c}(T,T')$~\cite{FP95}. In the limit
  $\varepsilon\rightarrow0^+$, in which typical simulations and experiments are
  carried out, the high-overlap glass states are only metastable. The timescale
  to escape from these metastable states can however be made extremely long,
  and vestiges of the Gardner transition can thus be well
  preserved.}\BibitemShut {Stop}%
\bibitem [{foo({\natexlab{b}})}]{footnote_coupling}%
  \BibitemOpen
  \href@noop {} {} {Note that cubic
  couplings are sometimes also denoted as $g_1\equiv g^{\mathrm{II}}$ and
  $g_2\equiv g^{\mathrm{I}}$, with possible differences in overall factors of
  order one.}\BibitemShut {Stop}%
\bibitem [{foo({\natexlab{c}})}]{footnote_Lagrange}%
  \BibitemOpen
  \href@noop {} {}  {We can derive the
  equations of motion by introducing the Lagrange multiplier
  field~\cite{Yaida16}, imposing the replicon constraints (see
  Appendix~\ref{Bresum3}).}\BibitemShut {Stop}%
\end{thebibliography}

\end{document}